\documentclass[11pt,a4paper]{article}

\usepackage[left=2.35cm,top=3cm,bottom=3cm,right=2.3cm]{geometry}

\usepackage{graphicx}
\usepackage[svgnames,x11names]{xcolor}
\definecolor{lightgray}{HTML}{D5D5D5}
\definecolor{lightergray}{HTML}{EEEEEE}
\usepackage{upgreek}
\usepackage{cite}
\usepackage{xcolor,colortbl}
\usepackage{slashed}

\definecolor{col}{rgb}{.4,.4,1}
\definecolor{col1}{rgb}{.4,.4,1}
\definecolor{col2}{rgb}{.4,.5,1}
\definecolor{col3}{rgb}{.4,.6,1}
\definecolor{col4}{rgb}{.4,.7,1}
\definecolor{hard}{rgb}{0.56,0.69,0.19}
\definecolor{stuff}{rgb}{0.88,0.88,0.88}
\definecolor{soft}{rgb}{0.88,0.61,0.14}
\definecolor{math1}{rgb}{0.37,0.51,0.71}
\definecolor{math2}{rgb}{0.88,0.61,0.14}
\definecolor{math3}{rgb}{0.56,0.69,0.19}
\definecolor{math4}{rgb}{0.92,0.39,0.21}
\definecolor{math5}{rgb}{0.53,0.47,0.70}
\definecolor{charge}{rgb}{0.8 0.15 0.15}

\usepackage{amsmath}
\usepackage{mathtools}
\usepackage{amsfonts}
\usepackage{amssymb}
\usepackage{slashed}
\usepackage{enumerate}
\usepackage[caption=false]{subfig}
\usepackage[colorlinks=true
,urlcolor=blue
,anchorcolor=blue
,citecolor=blue
,filecolor=blue
,linkcolor=blue
,menucolor=blue
,linktocpage=true
,pdfproducer=medialab
]{hyperref}
\numberwithin{equation}{section}

\usepackage{listings}
\usepackage{pgf}

\usepackage{xspace}

\usepackage{tikz}
\usetikzlibrary{shapes.geometric}
\usetikzlibrary{decorations.pathmorphing,positioning,decorations.pathreplacing,shapes,calc}
\usetikzlibrary{decorations.pathreplacing}
\usetikzlibrary{decorations.markings}
\usetikzlibrary{arrows}
\tikzset{
    photon/.style={decorate, decoration={snake,amplitude=1.5pt,segment length=4pt}},
    tightphoton/.style={decorate, decoration={snake,amplitude=1.5pt,segment length=5pt}},
    zigzag it/.style={decorate, decoration=zigzag},
    gluon/.style={decorate, draw=black,decoration={coil,amplitude=4pt, segment length=5pt}},
    tightgluon/.style={decorate, draw=black,decoration={coil,amplitude=2pt, segment length=3pt}}
}
\def\centerarc[#1](#2,#3)(#4:#5:#6)
    { \draw[#1] ({#2+#6*cos(#4)},{#3+#6*sin(#4)}) arc (#4:#5:#6); }

\newcommand\eik{\mathcal{E}}

\begin{document}
\thispagestyle{empty}

\begin{flushright}
FR-PHENO-2023-13
\end{flushright}
\vspace{3em}
\begin{center}
{\Large\bf Multiple soft-photon emission at next-to-leading power \\ to all orders}
\\
\vspace{3em}
{\sc
Tim\,Engel
}\\[2em]
{\sl Albert-Ludwigs-Universität Freiburg, Physikalisches Institut, \\
Hermann-Herder-Straße 3, D-79104 Freiburg, Germany}
\setcounter{footnote}{0}
\end{center}
\vspace{6ex}

\begin{center}
\begin{minipage}{15.3truecm}
{
This paper derives a next-to-leading power (NLP) soft theorem for multi-photon emission to all orders in the electromagnetic coupling constant, generalising the leading-power theorem of Yennie, Frautschi, and Suura. Working in the QED version of heavy-quark effective theory, multi-emission amplitudes are shown to reduce to single- and double-radiation contributions only. Single soft-photon emission, in turn, is described by the recent all-order extension of the Low-Burnett-Kroll theorem, where the tree-level formula is supplemented with a one-loop exact soft function. The same approach is used in this article to prove that the genuine double-emission contribution is tree-level exact. As a validation and a first non-trivial application of the multi-photon theorem, the real-real-virtual electron-line corrections to muon-electron scattering are calculated at NLP in the soft limit.
}
\end{minipage}
\end{center}

\newpage

\section{Introduction} \label{sec:intro}

The study of scattering amplitudes in the limit where external states become soft is of crucial importance both from a theoretical point of view and for precision phenomenology. As a source of infrared (IR) divergences, soft emissions complicate the calculation of cross sections and give rise to potentially large logarithms that spoil the convergence of the perturbative expansion. On the other hand, scattering amplitudes simplify drastically in the soft limit and universal factorised structures emerge. These factorisation theorems form the basis of IR subtraction schemes and the resummation of soft logarithms.

The soft structure of QED amplitudes is particularly simple and has long been fully understood at leading power (LP). The corresponding expansion is performed in terms of a power-counting parameter $\lambda$ after the common rescaling of the soft-photon momenta $k_a \to \lambda k_a$. For an amplitude (squared) with $n$ soft photons, the LP term in this expansion is given by the contributions that scale as $\lambda^{-n}$ ($\lambda^{-2n}$). In their seminal work~\cite{Yennie:1961ad}, Yennie, Frautschi, and Suura (YFS) showed that this LP term reduces to universal tree-level exact eikonal factors times the corresponding non-radiative amplitude to all orders and for an arbitrary number of soft-photon emissions. This is different in QCD where genuine loop corrections to the soft current exist~\cite{Catani:2000pi}. Together with the KLN theorem~\cite{Bloch:1937pw,Kinoshita:1962ur,Lee:1964is}, the YFS limit implies that soft virtual singularities in QED exponentiate. This simple structure significantly facilitates the construction of IR subtraction schemes. Moreover, it allows for the computation of an arbitrary number of soft-photon emissions and thereby for the resummation of the corresponding large logarithms.

Until recently, much less has been known about the all-order structure of single soft-photon emission at next-to-leading power (NLP). For an amplitude (squared) with $n$ soft photons this corresponds to the term that scales as $\lambda^{-n+1}$ ($\lambda^{-2n+1}$) in the power counting. At tree level, it was proven a long time ago by Low, Burnett, and Kroll (LBK)~\cite{Low:1958sn,Burnett:1967km} that not only the LP but also the NLP term in the soft expansion is fully determined by the non-radiative amplitude by means of a differential operator. This tree-level relation was later extended to polarised fermions~\cite{Tarasov:1968zrq,Fearing:1973eh,Kollatzsch:2022bqa}. Beyond tree level, additional contributions arise due to soft and collinear virtual corrections, depending on whether massive or massless fermions are considered. 

In the case of massless fermions, virtual corrections were studied already some time ago in~\cite{DelDuca:1990gz}, where radiative jet functions are introduced to take into account collinear effects. In recent years, there has been significant efforts to extend this massless version of the LBK theorem to QCD both in the framework of diagrammatic factorisation~\cite{Luo:2014wea,Bonocore:2015esa,Bonocore:2016awd,Laenen:2020nrt,vanBeekveld:2023gio,Pal:2023vec,Agarwal:2023fdk} as well as in soft-collinear effective theory~\cite{Larkoski:2014bxa,Beneke:2019oqx,Liu:2021mac}. Very recently, a complete generalisation to one-loop QCD amplitudes for arbitary processes was achieved in~\cite{Czakon:2023tld}. Analogous soft theorems have also been studied in gravity where even at next-to-next-to-leading power a relationship to the non-radiative process exists~\cite{Cachazo:2014fwa,Bern:2014vva,Beneke:2021umj}.

For massive fermions, the LBK theorem in QED was only recently generalised to one loop in~\cite{Engel:2021ccn} and to all orders in~\cite{Engel:2023ifn}. Since the leptons are taken to be heavy in this case, no collinear scale exists and radiative jet functions do not enter the theorem. Nevertheless, there are non-trivial corrections to the tree-level formula through soft virtual corrections, which were shown to be one-loop exact in~\cite{Engel:2023ifn}. This generalisation is particularly relevant in the context of fully-differential higher-order QED computations, where the NLP soft approximation is used to stabilise the numerical evaluation of real-emission amplitudes (next-to-soft stabilisation)~\cite{Banerjee:2021mty}. 

This article presents the extension of the all-order LBK theorem for massive fermions to an arbitrary number of soft-photon emissions. As in~\cite{Engel:2023ifn} this is done in the framework of the Abelian (QED) version of heavy-quark effective theory (HQET)~\cite{Eichten:1989zv,Georgi:1990um,Neubert:1993mb,Manohar:2000dt,Grozin:2004yc}, called heavy-lepton effective theory (HLET) in the following. To NLP, multi-radiation amplitudes are shown to be completely reducible to single- and double-emission contributions. While the single-emission term was already considered in~\cite{Engel:2023ifn}, the same approach is used here to show that the genuine double-emission contribution is tree-level exact. This generalises the YFS soft limit to NLP and opens the door for many applications, such as next-to-soft stabilisation of multi-radiation amplitudes and the extension of YFS resummation to NLP.

The paper is structured as follows. Section~\ref{sec:theorems} provides an overview on the NLP soft theorems for massive QED amplitudes. This includes the single-photon LBK theorem as well as the main result of this article given in~\eqref{eq:lbk_ng_allorder} for multiple emissions. The original tree-level proof of the LBK theorem is summarised at the beginning of Section~\ref{sec:qed} and extended to double- and triple-photon radiation in the following. By working in HLET, Section~\ref{sec:hlet} generalises these results to an arbitrary number of soft photons and to all orders. This follows closely the all-order single-emission proof of~\cite{Engel:2023ifn}, to which we refer for further details. A first, non-trivial application and validation of the derived NLP soft theorem is presented in Section~\ref{sec:validation} for muon-electron scattering. Finally, a conclusion and outlook is given in Section~\ref{sec:conclusion}.

\section{The LBK theorem and its generalisations} \label{sec:theorems}

The LBK theorem applies to the generic QED process
\begin{align} \label{eq:process_1g}
    \sum_{i=1}^m f_i(p_i) \to \gamma(k) \, ,
\end{align}
with $m$ fermions $f_i$ conventionally defined as incoming and a single emitted soft photon $\gamma$ as outgoing. The corresponding momenta are denoted by $p_i$ and $k$, respectively. Hard photons do not participate in soft interactions and only enter the hard matching coefficients. Hence, they do not have to be considered explicitly. The photon momentum $k$ is taken to be much smaller than all other scales in the process including all fermion masses $m_i$. The scale hierarchy thus reads
\begin{align} \label{eq:hierarchy_1g}
    \lambda \sim \frac{k}{p_i} \sim \frac{k}{m_i} \ll 1 \, ,
\end{align}
where the book-keeping parameter $\lambda$ is introduced to facilitate the power counting. The original version of the LBK theorem, proven in~\cite{Low:1958sn,Burnett:1967km}, holds at tree level and states that the NLP soft expansion of the radiative amplitude is completely determined by the non-radiative process. The corresponding relation for the unpolarised squared amplitude is given by\footnote{The symbol $\mathcal{M}$ denotes squared amplitudes in this article, while $\mathcal{A}$ is used for the amplitude itself.}
\begin{align} \label{eq:lbk_tree}
    \mathcal{M}_{m+1}^{(0)}(\{p\},k)
    = \big( E(k) + D(k) \big) \mathcal{M}_m^{(0)}(\{p\})
      + \mathcal{O}(\lambda^0) \, ,
\end{align}
with the eikonal factor
\begin{align} \label{eq:eikonal_squared}
    E(k) 
    &= -\sum_{i,l} Q_i Q_l \frac{p_i \cdot p_l}{k \cdot p_i k \cdot p_l}
\end{align}
and the LBK differential operator
\begin{align} \label{eq:lbkop_squared}
    D(k) 
    &= \sum_{i,l} Q_i Q_l \frac{p_l \cdot D_i}{k \cdot p_l}
    \, , \qquad
    \mathcal{D}_i^\mu(k) 
    = \frac{p_i^\mu}{k \cdot p_i} k \cdot \frac{\partial}{\partial p_i}
    - \frac{\partial}{\partial p_{i,\mu}} \, .
\end{align}
The fermion charge is denoted by $Q_i$ with $Q_i=e<0$ for an incoming particle or an outgoing antiparticle and $Q_i=+e$ otherwise. The form~\eqref{eq:lbk_tree} of the LBK theorem is not well suited for explicit calculations since the non-radiative amplitude is evaluated with momenta that violate momentum conservation $\sum_i p_i= \mathcal{O}(\lambda)$. It is more convenient to rewrite~\eqref{eq:lbk_tree} in terms of kinematic invariants $s_L$ of the non-radiative process, collectively denoted by $\{s\}$ in the following.\footnote{An alternative approach well suited for numerical calculations is based on momentum shifts~\cite{DelDuca:2017twk,Bonocore:2021cbv,Balsach:2023ema}.} According to~\cite{Engel:2021ccn}, the LBK theorem then takes the form
\begin{align} \label{eq:lbk_tree_inv}
    \mathcal{M}_{m+1}^{(0)}(\{p\},k)
    = \big( E(k) + \tilde{D}(k) \big) \mathcal{M}_m^{(0)}(\{s\},\{m^2\})
      + \mathcal{O}(\lambda^0) \, ,
\end{align}
with the modified LBK operator
\begin{align} \label{eq:lbk_inv}
    \tilde{D}(k) 
    &= \sum_{i,l} Q_i Q_l \frac{p_l \cdot \tilde{D}_i}{k \cdot p_l}
    \, , \qquad
   \tilde{\mathcal{D}}_i^\mu(k)
    = \sum_L \Big( \frac{p_i^\mu}{k \cdot p_i} k \cdot \frac{\partial s_L}{\partial p_i}
    - \frac{\partial s_L}{\partial p_{i,\mu}} \Big)
    \frac{\partial}{\partial s_L} \, .
\end{align}
As emphasized in~\cite{Engel:2021ccn}, different parametrisations of the invariants $\{s\}=\{s(\{p\},\{m^2\})\}$ differ at the level of $\mathcal{O}(\lambda)$ as a consequence of the aforementioned momentum conservation violation. It is therefore crucial to use the same definition both in the evaluation of the non-radiative contribution as well as in the calculation of the derivatives $\partial s_L/\partial p_i^\mu$. 

The tree-level LBK theorem~\eqref{eq:lbk_tree_inv} was first extended to one loop in~\cite{Engel:2021ccn} with the method of regions~\cite{Beneke:1997zp} and later to all orders in~\cite{Engel:2023ifn} with HLET. It was proven that beyond tree level the NLP contribution is still completely determined by the non-radiative process for unpolarised scattering. In the case of polarised fermions, this property is broken by factorisable QED corrections of the emitting leg or, equivalently, by spin-flipping magnetic contributions in HLET. These problematic contributions cancel at the level of the unpolarised squared amplitude and the all-order LBK theorem takes the form
\begin{align} \label{eq:lbk_allorder}
    \mathcal{M}_{m+1}(\{p\},k)
    = \big( E(k) + \tilde{D}(k) + S^{(1)}(k) \big) \mathcal{M}_m(\{s\},\{m^2\})
      + \mathcal{O}(\lambda^0) \, .
\end{align}
Compared to the tree-level theorem~\eqref{eq:lbk_tree_inv}, the additional contribution $S^{(1)}$ takes into account soft virtual corrections. It is the main result of the all-order proof~\cite{Engel:2023ifn} that this soft contribution is one-loop exact. It is given by
\begin{align}
    S^{(1)}(k) = \sum_{l,i,j \neq i} Q_i^2 Q_j Q_l
    \Big( \frac{p_i \cdot p_l}{k \cdot p_i k \cdot p_l}
    - \frac{p_j \cdot p_l}{k \cdot p_j k \cdot p_l} \Big)
    2 \mathcal{S}^{(1)}(p_i,p_j,k) \, ,
\end{align}
with the one-loop exact function
\begin{align} \label{eq:softfunc}
    \mathcal{S}^{(1)}(p_i,p_j,k)
    = \frac{m_i^2 k \cdot p_j}{\big((p_i \cdot p_j)^2 - m_i^2 m_j^2 \big) k \cdot p_i}
    \Big( p_i \cdot p_j I_1(p_i,k) + m_j^2 k \cdot p_i I_2(p_i,p_j,k) \Big)
\end{align}
defined in terms of the two simple integrals
\begin{align}  
	I_1(p_i,k) \label{eq:integral1}
	&= i \mu^{2\epsilon} \int \frac{\mathrm{d}^d \ell}{(2\pi)^d}
	\frac{1}{[\ell^2+i0][\ell \cdot p_i - k \cdot p_i + i 0]} \, ,
	\\
	I_2(p_i,p_j,k) \label{eq:integral2}
	&= i \mu^{2\epsilon} \int \frac{\mathrm{d}^d \ell}{(2\pi)^d}
	\frac{1}{[\ell^2+i0][-\ell \cdot p_j+i 0][\ell \cdot p_i - k \cdot p_i + i 0]} \, .
\end{align}
The analytic results for $I_1$ and $I_2$ with exact $\epsilon$ dependence can be found in Appendix A of~\cite{Engel:2021ccn}. 

In the following, the all-order proof of the LBK theorem of~\cite{Engel:2023ifn} is extended to processes with an arbitrary number of soft-photon emissions
\begin{align} \label{eq:process_ng}
    \sum_{i=1}^m f_i(p_i) \to \sum_{a=1}^n \gamma_a(k_a) \, .
\end{align}
One possibility to define the scale hierarchy for this more general setup is to have an ordered set of soft-photon momenta
\begin{align}
    \lambda_1 \sim \frac{k_1}{p_i} \sim \frac{k_1}{m_i}
    \ll ... \ll  \lambda_n \sim \frac{k_n}{p_i} \sim \frac{k_n}{m_i}
    \ll 1 \, .
\end{align}
In this case, the LBK theorem~\eqref{eq:lbk_allorder} can be applied iteratively. Starting with the smallest scale $\lambda_1$, its NLP expansion can be written in terms of the squared amplitude without the photon with momentum $k_1$. The smallest scale in this amplitude is now $\lambda_2$ and the procedure can be repeated until the reduction to the non-radiative process
\begin{align}  \label{eq:lbk_ordered}
    \mathcal{M}_{m+n} 
    = \prod_{a=1}^n \big( 
        E_a(k_a) + \tilde{D}_a(k_a) + S_a^{(1)}(k_a)
    \big) \mathcal{M}_m(\{s\},\{m^2\}) + \mathcal{O}(\lambda_1^0 ... \lambda_n^0)
\end{align}
is reached. There are two subtleties in interpreting~\eqref{eq:lbk_ordered}. First, different sets of invariants have to be used for the different LBK formulas as they originate from processes with a different number of external states. This is emphasized with an additional subscript $a$. Second, the LBK differential operators $\tilde{D}_a(k_a)$ do not commute. The order of the product in~\eqref{eq:lbk_ordered} is fixed with the left-most term corresponding to the smallest scale $k_1$. The soft theorem~\eqref{eq:lbk_ordered} can thus be evaluated via an iterative application of the LBK theorem starting with the formula for the largest scale applied to the non-radiative squared amplitude.

A more interesting situation arises for the unordered hierarchy
\begin{align}
    \lambda \sim \frac{k_a}{p_i} \sim \frac{k_a}{m_i} \ll 1 \, ,
\end{align}
where the single-emission theorem cannot be straightforwardly applied and needs to be extended accordingly. The corresponding NLP soft theorem is the main result of this paper and reads
\begin{align} \label{eq:lbk_ng_allorder}
    \mathcal{M}_{m+n}(\{p\},\{k\}) 
    &= \Bigg[ \prod_{a=1}^n E(k_a) 
    + \sum_{a=1}^n \prod_{c \neq a} E(k_c) \big( \tilde{D}(k_a)
        + S^{(1)}(k_a) \big)
    \nonumber \\ & \qquad
    + \sum_{a,b=1}^n \prod_{c \neq a,b} E(k_c) G^{(0)}(k_a,k_b)
    \Bigg] \mathcal{M}_m(\{s\},\{m^2\})
    + \mathcal{O}(\lambda^{-2n+2}) \, ,
\end{align}
with the tree-level exact universal function
\begin{align} \label{eq:2g_newobject}
    G^{(0)}(k_a,k_b)
    &= \sum_{i,k,l} \frac{ Q_i^2 Q_k Q_l  
    }{k_a \cdot p_i\, k_b \cdot p_i (k_a \cdot p_i + k_b \cdot p_i) k_a \cdot p_k\, k_b \cdot p_l}
    \big(p_k \cdot p_i \, p_l \cdot p_i \, k_a \cdot k_b
    \nonumber \\
    & \qquad \quad
        + p_k \cdot p_l \, k_a \cdot p_i \, k_b \cdot p_i
        -  p_k \cdot p_i \, k_a \cdot p_l \, k_b \cdot p_i 
        - p_l \cdot p_i \, k_b \cdot p_k \, k_a \cdot p_i \big)\, .
\end{align}
Amplitudes with multiple soft-photon emissions thus reduce at NLP to single- and double-radiation contributions only. At tree level, this was already proven in~\cite{Laenen:2008gt,Laenen:2010uz}, where the more general case of soft-gluon emission in QCD was studied. In fact, the tree-level expression~\eqref{eq:2g_newobject} agrees with (6.34) of~\cite{Laenen:2010uz} after interference with the eikonal approximation. The soft theorem~\eqref{eq:lbk_ng_allorder} can thus be viewed as the all-order generalisation of this result in the Abelian case. The corresponding all-order derivation builds upon the methodology of the single-emission proof of~\cite{Engel:2023ifn} and is structured in the following way. Section~\ref{sec:qed} studies soft-photon emission in QED and proves~\eqref{eq:lbk_ng_allorder} at tree level for up to $n=3$ emissions. These results are first reproduced in Section~\ref{sec:hlet} in HLET and then generalised to $n$ emissions. The purpose of rederiving these known tree-level results is to provide a pedagocical introduction to the methodology that forms the basis for the all-order proof in Section~\ref{sec:hlet_nphoton_allorder}.

\section{Multiple soft-photon emission in QED at tree level} \label{sec:qed}

This section studies the tree-level emission of up to three soft photons in QED and derives the soft theorem~\eqref{eq:lbk_ng_allorder} for these cases. Section~\ref{sec:qed_1photon} considers single-photon emission and provides a concise summary of the tree-level proof of the LBK theorem. Section~\ref{sec:qed_2photon} and Section~\ref{sec:qed_3photon} extend this derivation to double- and triple-photon emission, respectively.

\subsection{Single-photon emission} \label{sec:qed_1photon}

In the following, a brief summary of the original tree-level proof of the LBK theorem from~\cite{Low:1958sn,Burnett:1967km} is presented. A more detailed exposition of the proof can be found in Section~3.1 of~\cite{Engel:2021ccn}. 

The single-emission amplitude receives contributions both from diagrams where the photon is emitted from an external leg as well as from an internal line. As will become clear in the following, it is useful to split the amplitude into these two diagram types even though (or precisely because) they are not separately gauge invariant. In terms of external and internal emissions the amplitude reads
\begin{equation}
	\mathcal{A}_{m+1}^{(0)}
    = \Big(\sum_i \mathcal{A}_{k_1;}^{i,(0)} \Big) + \mathcal{A}_{;k_1}^{(0)}
	=  \Bigg(\,  \sum_i \begin{tikzpicture}[scale=.8,baseline={(0,-.1)}]
        	   \input{figures/diag_ext_1g}
       	   \end{tikzpicture}
	  \, \Bigg)
	   +
	   \begin{tikzpicture}[scale=.8,baseline={(0,-.1)}]
        	   \input{figures/diag_int_1g}
       	   \end{tikzpicture} \, .
\end{equation}
The semicolon in the external- and internal-emission amplitudes, $\mathcal{A}_{k_1;}^{i,(0)}$ and $\mathcal{A}_{;k_1}^{(0)}$, distinguishes between external and internal emission, respectively. The photon momentum on the l.h.s.\ of the semicolon corresponds to a photon that is emitted from an external leg, while a r.h.s.\ momentum denotes internal emission. In the subsequent sections, this notation naturally generalises to mixed multi-emission diagrams where some photons are emitted externally and others internally. For simplicity, the semicolon is omitted in the following if all momenta are emitted externally, \textit{i.e.}\ for single emission from the $i$-th external leg $\mathcal{A}_{k_1}^{i,(0)}=\mathcal{A}_{k_1;}^{i,(0)}$. 

Stripping the $i$-th external spinor from the non-radiative amplitude
\begin{align} \label{eq:subamp_tree}
    \mathcal{A}^{(0)}_m(\{p\}) = \Gamma_i^{(0)}(\{p\}) u(p_i) \, ,
\end{align}
the external emission amplitude can be written as
\begin{align} \label{eq:1g_tree_external_interm}
	\mathcal{A}_{k_1}^{i,(0)}
    = \frac{Q_i \Gamma_i^{(0)}(\{p\}_i) (\slashed{p_i}-\slashed{k_1}+m)
       \slashed{\epsilon_1}u(p_i)}{-2 k_1 \cdot p_i} \, ,
\end{align}
with $\{p\}_i=\{p_1,...,p_i-k_1,...,p_m\}$ and the shorthand notation for the photon polarisation vector $\epsilon_1^\mu = \epsilon^\mu(k_1)$. The expression~\eqref{eq:1g_tree_external_interm} can be expanded as
\begin{align}
	\mathcal{A}_{k_1}^{i,(0)}
    = \eik_i(k_1) \mathcal{A}_m^{(0)}(\{p\})
	+ \mathcal{A}_{k_1,\mathrm{NLP}}^{i,(0)}
    + \mathcal{O}(\lambda) \, ,
\end{align}
with the amplitude-level (in contrast to~\eqref{eq:eikonal_squared}) eikonal factor
\begin{align} \label{eq:eikamp}
    \eik_i(k_1) = \frac{Q_i \epsilon_1 \cdot p_i}{-k_1 \cdot p_i}
\end{align}
and the NLP contribution
\begin{align} \label{eq:singleNLP}
    \mathcal{A}_{k_1,\mathrm{NLP}}^{i,(0)}
    = -\eik_i(k_1)\,  \Big( k_1 \cdot \frac{\partial}{\partial p_i} \Gamma_i^{(0)}(\{p\}) \Big)u(p_i)
	+ \frac{Q_i \Gamma_i^{(0)}(\{p\}) \slashed{k_1} \slashed{\epsilon_1}u(p_i)}{2 k_1 \cdot p_i} \, .
\end{align}
The explicit dependence of $\Gamma_i^{(0)}$ on the fermion momenta $\{p\}$ is dropped in the following for the sake of simplicity.

The crucial insight of the original LBK proof~\cite{Low:1958sn,Burnett:1967km} -- which was inspired by~\cite{Adler:1966gc} -- is that to NLP the internal emission contribution can be determined by gauge invariance in the following way. The Ward identity for the radiative amplitude
\begin{align}
    \mathcal{A}_{m+1}^{(0)} 
    = \epsilon_{1,\mu} \mathcal{A}_{m+1}^{(0),\mu}
    =  \Big( \sum_i  \epsilon_{1,\mu} \mathcal{A}_{k_1}^{i,(0),\mu}  \Big)
    + \epsilon_{1,\mu} \mathcal{A}_{;k_1}^{(0),\mu}
\end{align}
implies
\begin{align} \label{eq:wardid_contracted}
    k_1 \cdot \mathcal{A}_{;k_1}^{(0)}
    = - \sum_i Q_i \Big( k_1 \cdot  \frac{\partial}{\partial p_i} \Gamma_i^{(0)} \Big) u(p_i)
    + \mathcal{O}(\lambda^2) \, .
\end{align}
The LP contribution separately satisfies the Ward identity due to charge conservation $\sum_i Q_i = 0$ and the second term on the r.h.s.\ of \eqref{eq:singleNLP} vanishes because of $\slashed{k_1} \slashed{k_1} = k_1^2=0$. Since internal emission at tree level cannot give rise to the propagator structure $1/{k_1}$, $\mathcal{A}_{;k_1}^{(0)}$ is local in $k_1$ and
\begin{align}
    \mathcal{A}_{;k_1}^{(0),\mu}
    = \mathcal{A}_{;k_1}^{(0),\mu}\Big|_{k_1=0} + \mathcal{O}(\lambda) \, .
\end{align}
Hence, the relation~\eqref{eq:wardid_contracted} also holds at the uncontracted level, which yields for the internal emission amplitude
\begin{align} \label{eq:singleNLP_internal}
    \mathcal{A}_{;k_1}^{(0)} 
    = -\sum_i Q_i \Big( \epsilon_1 \cdot \frac{\partial}{\partial p_i}
      \Gamma_i^{(0)} \Big) u(p_i) + \mathcal{O}(\lambda) \, .
\end{align}
It is important to emphasize that, contrary to statements made in earlier literature (see \text{e.g.}\ \cite{Bern:2014vva}), locality of the internal emission contribution is not preserved beyond tree level. It is broken by factorisable corrections of the emitting leg as well as the soft momentum region of loop integrals. The original proof of the LBK theorem is therefore only valid at tree level.

The internal emission result~\eqref{eq:singleNLP_internal} combines with the first term of~\eqref{eq:singleNLP} to $\epsilon_1 \cdot \mathcal{D}_i(k_1)$ with the LBK differential operator $\mathcal{D}_i$ defined in~\eqref{eq:lbkop_squared}. Squaring the amplitude, summing over spins and polarisations, and using
\begin{align}
    \frac{(\slashed{p_i}+m) \slashed{\epsilon_1} \slashed{k_1}
        + \slashed{k_1} \slashed{\epsilon_1} (\slashed{p_i} + m)
    }{2 k_1 \cdot p_i}
    = \frac{\epsilon_1 \cdot p_i}{k_1 \cdot p_i} \slashed{k_1}-\slashed{\epsilon_1}
    = \epsilon_1 \cdot \mathcal{D}_i(\slashed{p_i}+m)
\end{align}
yields the tree-level LBK theorem~\eqref{eq:lbk_tree}.

\subsection{Double-photon emission} \label{sec:qed_2photon}

This section extends the tree-level LBK theorem to processes with two emitted soft photons. Following the proof of the previous section, the amplitude can again be split according to the number of external and internal emissions. In this case, however, purely internal emission only contributes at next-to-next-to-leading power, \textit{i.e}\ at $\mathcal{O}(\lambda^0)$, and the NLP split simplifies to
\begin{align}  \label{eq:2g_split_tree}
	\mathcal{A}_{m+2}^{(0)}
    &= \Big( \sum_{i,j}  \mathcal{A}_{k_1 k_2}^{ij,(0)} \Big)
       + \Big(\sum_i \mathcal{A}_{k_1;k_2}^{i,(0)}+ (k_1 \leftrightarrow k_2)\Big)
       + \mathcal{O}(\lambda^0)
    \nonumber \\
	&= \Bigg(\, \sum_{i,j} 
        \begin{tikzpicture}[scale=.8,baseline={(0,-.1)}]
        	   \input{figures/diag_ext_2g}
       	\end{tikzpicture} \, \Bigg)
	   +
     \Bigg(\, \sum_i
	   \begin{tikzpicture}[scale=.8,baseline={(0,-.1)}]
        	   \input{figures/diag_int_2g}
       	   \end{tikzpicture}
       + (k_1 \leftrightarrow k_2)
       \, \Bigg)
       + \mathcal{O}(\lambda^0) \, .
\end{align}
As already mentioned in the previous section, mixed amplitudes with one external and one internal photon are denoted by $\mathcal{A}_{k_a;k_b}^{i,(0)}$, where the momentum on the r.h.s.\ of the semicolon refers to the photon that is emitted from an internal line. For simplicity, the semicolon is omitted in the purely external emission amplitude $\mathcal{A}_{k_1 k_2}^{ij,(0)}=\mathcal{A}_{k_1 k_2;}^{ij,(0)}$. The mixed diagrams scale as $\sim \lambda^{-1}$ and do not contribute at LP. Hence, to NLP it is sufficient to use the eikonal approximation for the externally emitted photon and the mixed amplitudes reduce to single-emission terms as
\begin{align} \label{eq:2g_int}
    \mathcal{A}_{k_1;k_2}^{i,(0)}
    = \eik_i(k_1) \mathcal{A}_{;k_2}^{(0)}
    + \mathcal{O}(\lambda^0) \, .
\end{align}
This also holds for the purely external emission amplitudes $\mathcal{A}_{k_1 k_2}^{ij,(0)}$ if the two photons are emitted from different legs, \textit{i.e.}\ for $i \neq j$. In this case, a NLP contribution of one leg always combines with a LP eikonal term of the other one, which yields
\begin{align} \label{eq:2g_ext_trivial}
    \mathcal{A}_{k_1 k_2}^{i\neq j,(0)}
    = \eik_i(k_1) \eik_j(k_2) \mathcal{A}_m^{(0)}
    + \eik_i(k_1) \mathcal{A}_{k_2,\mathrm{NLP}}^{j,(0)}
    + \eik_j(k_2) \mathcal{A}_{k_1,\mathrm{NLP}}^{i,(0)}
    + \mathcal{O}(\lambda^0)\, .
\end{align}

The only non-trivial amplitude to be calculated is the one for double emission from the same external leg given by
\begin{align} \label{eq:qed_2g_external}
    \mathcal{A}_{k_1 k_2}^{ii,(0)}
    &= \begin{tikzpicture}[scale=.8,baseline={(0,-.1)}]
        	   \input{figures/diag_ext_2g_sameleg}
       	   \end{tikzpicture}
    + (k_1 \leftrightarrow k_2)
    \nonumber \\
    &= \frac{
        Q_i^2 \Gamma_i^{(0)}(\{p\}_i) (\slashed{p_i}-\slashed{k_1}-\slashed{k_2}+m)
       \slashed{\epsilon_2} (\slashed{p_i}-\slashed{k_1}+m) \slashed{\epsilon_1} u(p_i)
       }{4 k_1 \cdot p_i (k_1 \cdot p_i + k_2 \cdot p_i - k_1 \cdot k_2)}
       + (k_1 \leftrightarrow k_2) \, ,
\end{align}
with $\{p\}_i=\{p_1,...,p_i-k_1-k_2,...,p_m\}$. This expands to
\begin{align} \label{eq:qed_2g_external_exp}
    \mathcal{A}_{k_1 k_2}^{ii,(0)}
    &= \frac{Q_i^2 \epsilon_1 \cdot p_i \epsilon_2 \cdot p_i
         }{k_1 \cdot p_i (k_1 \cdot p_i + k_2 \cdot p_i)} 
         \Big( \Gamma_i^{(0)} - \big( (k_1 + k_2) \cdot \frac{\partial}{\partial p_i} \Gamma_i^{(0)} \Big) \Big) u(p_i)
     \nonumber \\ & \quad
     + \frac{Q_i^2 \epsilon_1 \cdot p_i \epsilon_2 \cdot p_i k_1 \cdot k_2
         }{k_1 \cdot p_i (k_1 \cdot p_i + k_2 \cdot p_i)^2} \mathcal{A}_m^{(0)}
    \nonumber \\ & \quad
     - \frac{ Q_i^2
         \Gamma_i^{(0)} (\slashed{k_1}+\slashed{k_2}) \slashed{\epsilon_2} u(p_i)\epsilon_1 \cdot p_i
         }{2 k_1 \cdot p_i (k_1 \cdot p_i + k_2 \cdot p_i)}
     - \frac{Q_i^2
           \Gamma_i^{(0)} (\slashed{p_i}+m) \slashed{\epsilon_2} \slashed{k_1} 
           \slashed{\epsilon_1} u(p_i)
           }{4 k_1 \cdot p_i (k_1 \cdot p_i + k_2 \cdot p_i)}
    \nonumber \\ & \quad
     + (k_1 \leftrightarrow k_2) + \mathcal{O}(\lambda^0) \, .
\end{align}
The first line of~\eqref{eq:qed_2g_external_exp} takes into account the expansion of the non-radiative sub-amplitude $\Gamma_i(\{p\}_i)$. As in the previous section, the functional dependence on the external fermion momenta $\{p\}$ after the expansion is not explicitly displayed. The second and third line of~\eqref{eq:qed_2g_external_exp} correspond to the NLP contributions of the denominator and numerator in~\eqref{eq:qed_2g_external}, respectively. After writing the Dirac string of the second term in the third line of~\eqref{eq:qed_2g_external_exp} as
\begin{align}
    (\slashed{p_i}+m) \slashed{\epsilon_2} \slashed{k_1} \slashed{\epsilon_1} u(p_i)
    &= \big( \slashed{k_1} \slashed{\epsilon_1} 2  \epsilon_2 \cdot p_i 
    - \slashed{\epsilon_2} \slashed{\epsilon_1} 2 k_1 \cdot p_i
    - \slashed{k_1} \slashed{\epsilon_2} 2 \epsilon_1 \cdot p_i
    + 4 \epsilon_1 \cdot p_i \epsilon_2 \cdot k_1 \big) u(p_i)  \, ,
\end{align}
the expansion takes the form
\begin{align} \label{eq:qed_2g_external_exp_rewritten}
    \mathcal{A}_{k_1 k_2}^{ii,(0)}
    &= \frac{Q_i^2 \epsilon_1 \cdot p_i \epsilon_2 \cdot p_i
         }{k_1 \cdot p_i (k_1 \cdot p_i + k_2 \cdot p_i)} \mathcal{A}_m^{(0)}
    \nonumber \\ & \quad
    - \frac{Q_i^2}{k_1 \cdot p_i (k_1 \cdot p_i + k_2 \cdot p_i)}
    \Big(\epsilon_2 \cdot p_i \epsilon_1 \cdot p_i  \big( k_1 \cdot \frac{\partial}{\partial p_i} \Gamma_i^{(0)} \big)
     + \epsilon_2 \cdot p_i \frac{\Gamma_i^{(0)} \slashed{k_1} \slashed{\epsilon_1}}{2} 
     + (k_1 \leftrightarrow k_2) \Big) u(p_i)
    \nonumber \\ & \quad
    + \frac{Q_i^2 \epsilon_1 \cdot p_i \epsilon_2 \cdot p_i k_1 \cdot k_2 
        }{k_1 \cdot p_i (k_1 \cdot p_i + k_2 \cdot p_i)^2} \mathcal{A}_m^{(0)}
    -  \frac{Q_i^2 \epsilon_1 \cdot p_i \epsilon_2 \cdot k_1 
    }{k_1 \cdot p_i (k_1 \cdot p_i + k_2 \cdot p_i)} \mathcal{A}_m^{(0)}
    + \frac{Q_i^2\Gamma_i^{(0)} \slashed{\epsilon_2} \slashed{\epsilon_1} u(p_i)
        }{2(k_1 \cdot p_i + k_2 \cdot p_i)}
    \nonumber \\ & \quad
    + (k_1 \leftrightarrow k_2) + \mathcal{O}(\lambda^0) \, .
\end{align}
The second line of~\eqref{eq:qed_2g_external_exp_rewritten} reduces to the single-emission terms $\eik_i(k_2)\mathcal{A}_{k_1,\text{NLP}}^{i,(0)}$ and $\eik_i(k_1)\mathcal{A}_{k_2,\text{NLP}}^{i,(0)}$, with the NLP contribution defined in~\eqref{eq:singleNLP}. This follows after adding the corresponding term with $(k_1 \leftrightarrow k_2)$ and applying the partial fraction (eikonal) identity
\begin{align} \label{eq:partial_fraction}
    \frac{1}{k_1 \cdot p_i (k_1 \cdot p_i + k_2 \cdot p_i)} + (k_1 \leftrightarrow k_2)
    = \frac{1}{k_1 \cdot p_i k_2 \cdot p_i} \, .
\end{align}
The third line gives rise to a new structure which is also proportional to the non-radiative amplitude. The final result reads
\begin{align} \label{eq:2g_ext_nontrivial}
    \mathcal{A}_{k_1 k_2}^{ii,(0)}
    & = \eik_i(k_1) \eik_i(k_2) \mathcal{A}_m^{(0)}
        + \eik_i(k_2) \mathcal{A}_{k_1,\mathrm{NLP}}^{i,(0)}
        +\eik_i(k_1) \mathcal{A}_{k_2,\mathrm{NLP}}^{i,(0)}
        + \mathcal{G}_i^{(0)}(k_1,k_2) \mathcal{A}_m^{(0)}
        + \mathcal{O}(\lambda^0) \, ,
\end{align}
with the new term given by
\begin{align} \label{eq:eiknts_amp}
    \mathcal{G}_i^{(0)}(k_1,k_2)
     = Q_i^2 \frac{
         \epsilon_1 \cdot p_i \, \epsilon_2 \cdot p_i \, k_1 \cdot k_2
         + \epsilon_1 \cdot \epsilon_2 \, k_1 \cdot p_i \, k_2 \cdot p_i
         - \epsilon_1 \cdot p_i \, \epsilon_2 \cdot k_1 \, k_2 \cdot p_i
         - \epsilon_2 \cdot p_i \, \epsilon_1 \cdot k_2 \, k_1 \cdot p_i
     }{k_1 \cdot p_i \, k_2 \cdot p_i (k_1 \cdot p_i + k_2 \cdot p_i)} \, .
\end{align}
Combining \eqref{eq:2g_int}, \eqref{eq:2g_ext_trivial}, and \eqref{eq:2g_ext_nontrivial} gives for the total NLP amplitude
\begin{align} \label{eq:lbk_2g_tree_amp}
    \mathcal{A}_{m+2}^{(0)}
    &= \eik(k_1) \eik(k_2) \mathcal{A}_m^{(0)}
        + \eik(k_2) \mathcal{A}_{k_1,\text{NLP}}^{(0)}
        + \eik(k_1) \mathcal{A}_{k_2,\text{NLP}}^{(0)}
        + \mathcal{G}^{(0)}(k_1,k_2) \mathcal{A}_m^{(0)}
        + \mathcal{O}(\lambda^0) \, ,
\end{align}
with
\begin{align} \label{eq:eik_sum}
    \eik(k_a) = \sum_i \eik_i(k_a), \qquad
    \mathcal{G}^{(0)}(k_a,k_b) = \sum_i \mathcal{G}^{(0)}_i(k_a,k_b)\, ,
\end{align}
and the total NLP contribution to the single-emission amplitude given by
\begin{align} \label{eq:single_nlp}
    \mathcal{A}_{k_a,\text{NLP}}^{(0)}
    =  \Big( \sum_i \mathcal{A}_{k_a,\text{NLP}}^{i,(0)} \Big)
        + \mathcal{A}_{;k_a}^{(0)} \, .
\end{align}
The single-emission results of the previous section~\eqref{eq:singleNLP} and~\eqref{eq:singleNLP_internal} together with~\eqref{eq:lbk_2g_tree_amp} implies for the unpolarised squared amplitude
\begin{align} \label{eq:lbk_2g_tree}
    \mathcal{M}_{m+2}^{(0)}(\{p\},\{k_1,k_2\})
    &= \Big( E(k_1) E(k_2) + E(k_1) \tilde{D}(k_2) + E(k_2) \tilde{D}(k_1)
    \nonumber \\ & \quad
    + 2 G^{(0)}(k_1,k_2) \Big) \mathcal{M}_m^{(0)}(\{s\},\{m^2\})
    + \mathcal{O}(\lambda^{-2})\, ,
\end{align}
with $E$, $D$, and $G^{(0)}$ defined in~\eqref{eq:eikonal_squared}, \eqref{eq:lbkop_squared}, and $\eqref{eq:2g_newobject}$, respectively. As the last step, the factor of $2$ in~\eqref{eq:lbk_2g_tree} from the interference can be rewritten as
\begin{align}
    2 G^{(0)}(k_1,k_2) = G^{(0)}(k_1,k_2) + G^{(0)}(k_2,k_1) \, ,
\end{align}
which proves the soft theorem~\eqref{eq:lbk_ng_allorder} at tree level for $n=2$.

\subsection{Triple-photon emission} \label{sec:qed_3photon}

The extension to the three-photon case proceeds along the same line as in the previous sections. The split of the amplitude into internal and external emissions reads
\begin{align}
	\mathcal{A}_{m+3}^{(0)}
    &= \Big( \sum_{i,j,l} \mathcal{A}_{k_1 k_2 k_3}^{ijl,(0)} \Big)
       + \Big(\sum_{i,j} \mathcal{A}_{k_1 k_2;k_3}^{ij,(0)} 
       + (k_{1,2} \leftrightarrow k_3)\Big)
       + \mathcal{O}(\lambda^{-1})
    \nonumber \\
	&= \Bigg( \, \sum_{i,j,l} 
        \begin{tikzpicture}[scale=.8,baseline={(0,-.1)}]
        	   \input{figures/diag_ext_3g}
       	\end{tikzpicture}\, \Bigg)
	   +
     \Bigg(\, \sum_{i,j}
	   \begin{tikzpicture}[scale=.8,baseline={(0,-.1)}]
        	   \input{figures/diag_int_3g}
       	   \end{tikzpicture}
       + (k_{1,2} \leftrightarrow k_3)
       \, \Bigg)
       + \mathcal{O}(\lambda^{-1}) \, ,
\end{align}
with the shorthand notation 
\begin{align} \label{eq:2perm_short}
    (k_{1,2} \leftrightarrow k_3) = (k_1 \leftrightarrow k_3) + (k_2 \leftrightarrow k_3) \, .
\end{align}
In complete analogy to~\eqref{eq:2g_int} and \eqref{eq:2g_ext_trivial}, internal emission as well as external radiation from different legs reduces to single-emission terms. This yields for the corresponding amplitudes
\begin{align}
    \mathcal{A}_{k_1 k_2;k_3}^{ij,(0)}
    = \eik_i(k_1) \eik_j(k_2) \mathcal{A}_{;k_3}^{(0)}
    + \mathcal{O}(\lambda^{-1})
\end{align}
and
\begin{align}
    \mathcal{A}_{k_1 k_2 k_3}^{i \neq j \neq l,(0)}
    &= \eik_i(k_1) \eik_j(k_2) \eik_l(k_3) \mathcal{A}_m^{(0)}
    \nonumber \\ & \quad
    + \eik_i(k_1) \eik_j(k_2) \mathcal{A}_{k_3,\mathrm{NLP}}^{l,(0)}
    + \eik_i(k_1) \eik_l(k_3) \mathcal{A}_{k_2,\mathrm{NLP}}^{j,(0)}
    + \eik_j(k_2) \eik_l(k_3) \mathcal{A}_{k_1,\mathrm{NLP}}^{i,(0)}
    \nonumber \\ & \quad
    + \mathcal{O}(\lambda^{-1}) \, .
\end{align}
Similarly, in the case where two photons are emitted from the same leg, the external emission contribution can be written in terms of single- and double-emissions according to
\begin{align}
     \mathcal{A}_{k_1 k_2 k_3}^{ii \neq l,(0)}
     &= \eik_l(k_3) \mathcal{A}_{k_1 k_2}^{ii,(0)} 
     + \eik_i(k_1) \eik_i(k_2) \mathcal{A}_{k_3,\text{NLP}}^{l,(0)}
     + \mathcal{O}(\lambda^{-1})\, .
\end{align}
Inserting the expression~\eqref{eq:2g_ext_nontrivial} for the double-emission term then yields
\begin{align}
     \mathcal{A}_{k_1 k_2 k_3}^{ii \neq l,(0)}
     &= \eik_i(k_1) \eik_i(k_2) \eik_l(k_3) \mathcal{A}_m^{(0)}
    \nonumber \\ & \quad
    + \eik_i(k_1) \eik_i(k_2) \mathcal{A}_{k_3,\mathrm{NLP}}^{l,(0)}
    + \eik_i(k_1) \eik_l(k_3) \mathcal{A}_{k_2,\mathrm{NLP}}^{i,(0)}
    + \eik_i(k_2) \eik_l(k_3) \mathcal{A}_{k_1,\mathrm{NLP}}^{i,(0)}
    \nonumber \\ & \quad
    + \eik_l(k_3) \mathcal{G}_i^{(0)}(k_1,k_2) \mathcal{A}_m^{(0)}
    \nonumber \\ & \quad
    + \mathcal{O}(\lambda^{-1}) \, .
\end{align}

A new structure can only come from the same-leg triple emission contribution
\begin{align} \label{eq:qed_3g_crossedvertex}
    \mathcal{A}_{k_1 k_2 k_3}^{iii,(0)}
    &= \begin{tikzpicture}[scale=.8,baseline={(0,-.1)}]
    \input{figures/diag_ext_3g_sameleg_perm}
       	   \end{tikzpicture}
    = \begin{tikzpicture}[scale=.8,baseline={(0,-.1)}]
        	   \input{figures/diag_ext_3g_sameleg}
       	   \end{tikzpicture}
    + \sum_\sigma \big(\sigma(k_1,k_2,k_3)\big) \, ,
\end{align}
where the crossed vertex denotes all possible attachments of the three photons. This corresponds to summing over all possible permutations $\sigma$ of the three photon momenta as indicated on the r.h.s.\ of~\eqref{eq:qed_3g_crossedvertex}. The calculation of the NLP expansion of~\eqref{eq:qed_3g_crossedvertex} is conceptually the same as for the double-emission case~\eqref{eq:qed_2g_external}. The result is
\begin{align} \label{eq:amp_nts_3g_sameleg}
    \mathcal{A}_{k_1 k_2 k_3}^{iii,(0)}
    & = \eik_i(k_1) \eik_i(k_2) \eik_i(k_3) \mathcal{A}_m^{(0)}
    \nonumber \\ & \quad
    + \eik_i(k_1) \eik_i(k_2) \mathcal{A}_{k_3,\mathrm{NLP}}^{i,(0)}
    + \eik_i(k_1) \eik_i(k_3) \mathcal{A}_{k_2,\mathrm{NLP}}^{i,(0)}
    + \eik_i(k_2) \eik_i(k_3) \mathcal{A}_{k_1,\mathrm{NLP}}^{i,(0)}
    \nonumber \\ & \quad
    + \eik_i(k_3) \mathcal{G}_i^{(0)}(k_1,k_2) \mathcal{A}_m^{(0)}
    + \eik_i(k_2) \mathcal{G}_i^{(0)}(k_1,k_3)\mathcal{A}_m^{(0)}
    +  \eik_i(k_1) \mathcal{G}_i^{(0)}(k_2,k_3)
     \mathcal{A}_m^{(0)}
     \nonumber \\ & \quad
    + \mathcal{O}(\lambda^{-1}) \, .
\end{align}
Also this case reduces to single- and double-emission objects without giving rise to additional structures. As a result, the triple-emission amplitude to NLP is given by
\begin{align} \label{eq:lbk_3g_tree_amp}
    \mathcal{A}_{m+3}^{(0)}
    &= \eik(k_1) \eik(k_2) \eik(k_3) \mathcal{A}_m^{(0)}
    \nonumber \\ & \quad
    + \Big( \eik(k_1) \eik(k_2)  \mathcal{A}_{k_3,\text{NLP}}^{(0)} 
    + \eik(k_3) \mathcal{G}^{(0)}(k_1,k_2) \mathcal{A}_m^{(0)}
    + (k_{1,2} \leftrightarrow k_3) \Big)
    + \mathcal{O}(\lambda^{-1}) \, ,
\end{align}
with $\eik$ and $\mathcal{G}^{(0)}$ defined in~\eqref{eq:eik_sum} and the NLP single-emission contribution $\mathcal{A}_{k_a,\text{NLP}}^{(0)}$ given in~\eqref{eq:single_nlp}. At the level of the unpolarised squared amplitude this implies the soft theorem~\eqref{eq:lbk_ng_allorder} at tree level for $n=3$.

While it might be possible to extend these calculations to an arbitrary number of soft-photon emissions, the simple emerging structure becomes much more transparent in HLET. This will be particularly important when taking into account loop corrections.

\section{Multiple soft-photon emission in HLET to all orders} \label{sec:hlet}

This section studies soft-photon emission in HLET. In this effective field theory the simple structure of the NLP soft limit is much more transparent than in QED. This allows for a generalisation of the previous results to an arbitrary number of photon emissions and to all orders. Section~\ref{sec:eft} introduces the HLET Lagrangian and its Feynman rules. Based on this, so-called eikonal identities are introduced in Section~\ref{sec:eikid}, which form the key component of the following calculations. As a first step, the tree-level results of Section~\ref{sec:qed} for single-, double-, and triple-photon emission are reproduced in Sections~\ref{sec:hlet_1photon}, \ref{sec:hlet_2photon}, and \ref{sec:hlet_3photon}, respectively. The generalisation to an arbitrary number of photons is presented in Section~\ref{sec:hlet_nphoton} and extended to all orders in Section~\ref{sec:hlet_nphoton_allorder}. This completes the proof of the soft theorem~\eqref{eq:lbk_ng_allorder}.

\subsection{Heavy-lepton effective theory} \label{sec:eft}

HLET is the Ablian version of HQET~\cite{Eichten:1989zv,Georgi:1990um,Neubert:1993mb,Manohar:2000dt,Grozin:2004yc} and describes interactions between heavy leptons and soft photons. The movement of the heavy leptons is only slightly affected by the soft interactions. As a result, the lepton momenta $p_j$ stay close to their mass shell and can be decomposed as
\begin{align}
    p_j^\mu = m_j v_j^\mu + q_j^\mu \, ,
\end{align}
with $v_j^2=1$ and $|q_j^2| \ll m_j^2$. The large component of the lepton field $\psi_j$ is then given by
\begin{align}
    h_{v_j}(x) = e^{i m_j v_j \cdot x} P_{+_j} \psi_j(x) \, ,
\end{align}
with the projection operator $P_{+_j}=(1+\slashed{v_j})/2$. The HLET Lagrangian is obtained by integrating out the heavy degrees of freedom and by expanding in the small parameter $\lambda \sim |q_j^2|/m_j^2 \ll 1$. This yields to NLP~\cite{Eichten:1990vp,Falk:1990pz}
\begin{align}
	\label{eq:lag_hqet}
	\mathcal{L}_\text{HLET}
	 = \mathcal{L}_\text{LP} 
	 + \mathcal{L}_\text{kin} + \mathcal{L}_\text{mag}
	 + \mathcal{O}(\lambda^2) \, ,
\end{align}
with the LP term
\begin{align}
	\label{eq:lag_lp}
	\mathcal{L}_\text{LP} 
	& = \sum_{j=1}^m \Big( \bar{h}_{v_j} i v_j \cdot D\, h_{v_j} \Big)-  \frac{1}{4}F_{\mu\nu} F^{\mu\nu}
\end{align}
and the subleading kinetic and magnetic parts
\begin{align}
	 \mathcal{L}_\text{kin} 
	 & = \sum_{j=1}^m \frac{1}{2 m_j} \bar{h}_{v_j} (i D_{\perp_j})^2 h_{v_j} \, ,
	 \\
	 \label{eq:lag_mag}
	  \mathcal{L}_\text{mag}
	  & = \sum_{j=1}^m \frac{e C_\text{mag}}{4m_j} \bar{h}_{v_j} \sigma_{\mu\nu} F^{\mu\nu} h_{v_j} \, .
\end{align}
The usual definitions are used for the covariant derivative $D^\mu = \partial^\mu - i e A^\mu$, the electromagnetic field strength $F^{\mu\nu} = \partial^\mu A^\nu - \partial^\nu A^\mu$, and $\sigma_{\mu\nu}=i[\gamma^\mu,\gamma^\nu]/2$. Furthermore, the perpendicular component of the metric w.r.t.\ the direction $v_j$ is denoted by
\begin{align}
	g_{\perp_j}^{\mu\nu} = g^{\mu\nu}-v_j^\mu v_j^\nu \, ,
\end{align}
which implies
\begin{align} \label{eq:perpcomp}
	p_{\perp_j}^\mu
	= g_{\perp_j}^{\mu \nu} p_\nu
	= p^\mu - v_j^\mu v_j \cdot p
\end{align}
for any four-momentum $p$. The Wilson coefficient of the kinetic NLP Lagrangian $\mathcal{L}_\text{kin}$ does not receive loop corrections as a consequence of reparametrisation invariance under small $\mathcal{O}(\lambda)$ changes of the directions $v_j$~\cite{Luke:1992cs,Chen:1993np}. The non-trivial Wilson coefficient $C_\text{mag}$ is known to three loops~\cite{Czarnecki:1997dz,Grozin:2007fh}.

The Feynman rules for the Lagrangian~\eqref{eq:lag_hqet} are listed in the following. All labelled momenta are taken to be incoming and the fermion charge $Q_j$ is defined as in Section~\ref{sec:theorems}. The LP Lagrangian yields the familiar eikonal Feynman rules
\begin{align}
\label{eq:eikonal_rules}
	\begin{tikzpicture}[scale=.8,baseline={(0,-.1)}]

   \draw[line width=.3mm,style=double]  (-1,0) -- (1,0);
   \node at (-.8,.3) {\footnotesize{$v_j,q$}};

       \end{tikzpicture}
       = \frac{i}{q \cdot v_j + i 0} P_{+_j}
       \, , \quad\quad\quad
       	\begin{tikzpicture}[scale=.8,baseline={(0,-.1)}]

   \draw[line width=.3mm,style=double]  (-1,0) -- (1,0);
   \draw[line width=.3mm, photon] (0,0) -- (0,1) node[right] {\footnotesize{$k,\mu$}};

   \node at (-.8,.3) {\footnotesize{$v_j,q$}};

       \end{tikzpicture}
       = -i Q_j v_j^\mu \, .
\end{align}
Since all fermions are taken to be incoming in~\eqref{eq:process_ng}, the arrows denoting the charge flow are not explicitly displayed (as for the QED diagrams in Section~\ref{sec:qed}). In contrast to QED, HLET leptons are represented with a double line. At NLP, the kinetic term $\mathcal{L}_\text{kin}$ gives rise to the three vertices
\begin{align}
\label{eq:vertex_kin}
	\begin{tikzpicture}[scale=.8,baseline={(0,-.1)}]

   \draw[line width=.3mm,style=double]  (-1,0) -- (1,0);
   \node at (-.8,.3) {\footnotesize{$v_j,q$}};
   \node[rectangle,fill=black] at (0,0) {};

      	\end{tikzpicture}
       	=  \frac{i}{2m_j} q_{\perp_j}^2
	\, , \quad\quad
	\begin{tikzpicture}[scale=.8,baseline={(0,-.1)}]

   \draw[line width=.3mm,style=double]  (-1,0) -- (1,0);
   \node at (-.8,.3) {\footnotesize{$v_j,q$}};
   \node[rectangle,fill=black] at (0,0) {};
   \draw[line width=.3mm, photon] (0,0) -- (0,1) node[right] {\footnotesize{$k,\mu$}};

      	\end{tikzpicture}
	= \frac{-i Q_j}{2m_j} (2q+k)^\mu_{\perp_j}
	\, , \quad\quad
	\begin{tikzpicture}[scale=.8,baseline={(0,-.1)}]

   \draw[line width=.3mm,style=double]  (-1,0) -- (1,0);
   \node[rectangle,fill=black] at (0,0) {};
   \node at (-.8,.3) {\footnotesize{$v_j,q$}};
   \draw[line width=.3mm, photon] (0,0) -- (-.6,1) node[left]{\footnotesize{$\mu$}};
   \draw[line width=.3mm, photon] (0,0) -- (.6,1) node[right]{\footnotesize{$\nu$}};

      	\end{tikzpicture}
	= \frac{i Q_j^2}{m_j} g_{\perp_j}^{\mu\nu} \, .
\end{align}
The only vertex with a non-trivial Dirac structure comes from the magnetic term  $\mathcal{L}_\text{mag}$ and reads
\begin{align}
\label{eq:vertex_mag}
	\begin{tikzpicture}[scale=.8,baseline={(0,-.1)}]

   \draw[line width=.3mm,style=double]  (-1,0) -- (1,0);
   \node at (-.8,.3) {\footnotesize{$v_j,q$}};
   \node[draw,isosceles triangle,fill=black,rotate=-90,minimum size=.3cm,inner sep = 0pt]{};
   \draw[line width=.3mm, photon] (0,0) -- (0,1) node[right] {\footnotesize{$k,\mu$}};

      	\end{tikzpicture}
       	=  \frac{Q_j C_\text{mag}}{2m_j} \sigma_{\mu\nu} k^\nu \, .
\end{align}

In order to describe the scattering process~\eqref{eq:process_ng} with HLET, 
%
%
the Lagrangian~\eqref{eq:lag_hqet} has to be supplemented with the most general, gauge-invariant operators (external currents) that generate this process, \textit{i.e.}
\begin{align} \label{eq:lp_vertices}
	\mathcal{L} 
	= \mathcal{L}_\text{HLET}  
	+ \mathcal{O}_\text{LP} + \mathcal{O}_\text{NLP} \, .
\end{align}
The LP operator, $\mathcal{O}_\text{LP}$, consists of the $m$ fermion fields only. The NLP operator, $\mathcal{O}_\text{NLP}$, contains in addition one covariant derivative which raises the mass dimension of the operator by one unit. This gives rise to the following vertices:
\begin{align} \label{eq:ONLP_vertices}
	\begin{tikzpicture}[scale=1.,baseline={(0,-.1)}]
        		\input{figures/vertex_OLP}
      	\end{tikzpicture}
	\qquad\qquad\qquad
	\begin{tikzpicture}[scale=1.,baseline={(0,-.1)}]
        		\input{figures/vertex_ONLP1}
      	\end{tikzpicture}
	\qquad\qquad\qquad
	\begin{tikzpicture}[scale=1.,baseline={(0,-.1)}]
        		\input{figures/vertex_ONLP2}
      	\end{tikzpicture}
\end{align}
The exact structure of the above vertices does not matter in the following. The contribution of the first two vertices in the calculation of radiative amplitudes can be determined through the matching to the non-radiative process with off-shell external states. As for the single-emission amplitude~\eqref{eq:1g_tree_external_interm}, the QED spinors are evaluated with on-shell momenta that do not satisfy momentum conservation. Expanding the corresponding QED amplitude in the hard loop momentum region yields
\begin{align}
    \label{eq:nonrad_matching}
	\mathcal{A}_{\mathcal{O}_\text{LP}}(\{q\})
	&= \langle 0
	| \mathcal{O}_\text{LP}
	| \prod_{j=1}^m h_{v_j}(q_j) \rangle
    = \begin{tikzpicture}[scale=1.,baseline={(0,-.1)}]
        		\input{figures/vertex_OLP_offshell}
        \end{tikzpicture}
    = \mathcal{A}_{m}(\{p\}) \, ,
    \\
    \label{eq:nonrad_matching_photon}
	\mathcal{A}_{\mathcal{O}_\text{NLP}}(\{q\})
	& = \langle 0
	| \mathcal{O}_\text{NLP}
	| \prod_{j=1}^m h_{v_j}(q_j) \rangle
    = \begin{tikzpicture}[scale=1.,baseline={(0,-.1)}]
        		\input{figures/vertex_ONLP1_offshell}
        \end{tikzpicture}
    = \sum_{j=1}^m \Big( q_j \cdot \frac{\partial}{\partial p_j} \Gamma_j(\{p\}) \Big) u(p_j) \, ,
\end{align}
where the non-radiative QED (sub)amplitudes $\mathcal{A}_m=\Gamma_j u(p_j)$ are evaluated with on-shell momenta $p_j=m_j v_j$. In the case of the third vertex in~\eqref{eq:ONLP_vertices}, single-emission processes have to be considered for the matching due to the additional photon. This is discussed in Section~\ref{sec:hlet_1photon} in the context of the tree-level LBK theorem.

\subsection{Eikonal identities} \label{sec:eikid}

The all-order proof of the LBK theorem presented in~\cite{Engel:2023ifn} is based on an extension of the well-known on-shell eikonal identity to account for single-photon emission from off-shell lines. The identity is proven in Appendix A of~\cite{Engel:2023ifn}. While the on-shell identity does not depend on the number of external photons in the process, the off-shell version generalises non-trivially to multiple radiation. This section extends the single-emission identity accordingly. The occurrence of off-shell emissions originates from the NLP HLET vertices~\eqref{eq:vertex_kin} and~\eqref{eq:vertex_mag}. These vertices give rise to the generic diagrams~\eqref{eq:hlet_ngNLP_general_tree} at tree level and~\eqref{eq:hlet_ngNLP_general_allorder} at loop level. As shown in detail in the corresponding Sections~\ref{sec:hlet_nphoton} and~\ref{sec:hlet_nphoton_allorder}, the off-shell eikonal identity can be used to derive remarkably simple expressions for these amplitudes. 

The on-shell eikonal identity for $n$ incoming momenta $p_i$ is given by
\begin{equation}
\label{eq:eikid_conv}
         \mathcal{R}_n 
         = \begin{tikzpicture}[scale=.8,baseline={(0,-.1)}]
       	 \input{figures/eik_perm}
       	\end{tikzpicture}
    	= \prod_{i=1}^n \frac{Q v^{\mu_i}}{p_i \cdot v} \, ,
\end{equation}
where the crossed vertex represents all possible attachments of the photon lines as in~\eqref{eq:qed_3g_crossedvertex}. This simple structure emerges from various cancellations among the diagrams in 
\begin{align} \label{eq:eikid_conv_sum}
    \sum_{n_1+n_2=n-1}
        \begin{tikzpicture}[scale=.8,baseline={(0,-.1)}]
       	 \input{figures/eik_insert}
       	\end{tikzpicture}
    = \frac{Q v^{\mu_n}}{p_n \cdot v}
	\begin{tikzpicture}[scale=.8,baseline={(0,-.1)}]
        \input{figures/eik_noperm_less}
       \end{tikzpicture} \, .
\end{align}
As shown in detail in Appendix A.1 of~\cite{Engel:2023ifn}, all terms but one cancel after partial fraction decomposition of the two propagators to the left and to the right of the photon line with momentum $p_n$. This yields the relation
\begin{align} \label{eq:eikid_conv_recursion}
   \mathcal{R}_{n} = \frac{Q v^{\mu_n}}{p_n \cdot v} \mathcal{R}_{n-1} \, ,
\end{align}
which recursively applied proves~\eqref{eq:eikid_conv}.

The NLP vertices~\eqref{eq:vertex_kin} and \eqref{eq:vertex_mag} give rise to sub-diagrams of the form
\begin{align} \label{eq:eikid_gen_diag}
    	\mathcal{T}_{n,m}^{(q)}= \begin{tikzpicture}[scale=.8,baseline={(0,-.1)}]
       	 \input{figures/eikgen_perm}
       	\end{tikzpicture} \, ,
\end{align}
where the photons attach to a heavy-lepton line with residual momentum $q$. The small blob on the left side of the diagram indicates that the lepton is off shell. The momenta of the external photons are denoted by $k_a$, while $\ell_i$ is used for virtual photon lines. As shown in Appendix A.2 of~\cite{Engel:2023ifn}, the cancellation is less complete in this case and the relation~\eqref{eq:eikid_conv_sum} is modified to
\begin{align} \label{eq:eikid_gen_sum}
    \sum_{n_1+n_2=n-1} 
    \begin{tikzpicture}[scale=.8,baseline={(0,-.1)}]
       	 \input{figures/eikgen_insert}
       	\end{tikzpicture}
    =  \frac{-Q v^{\mu_n}}{p_n \cdot v}
	\Bigg(
        \begin{tikzpicture}[scale=.8,baseline={(0,-.1)}]
        \input{figures/eikgen_noperm_less2}
        \end{tikzpicture}
        -
	\begin{tikzpicture}[scale=.8,baseline={(0,-.1)}]
        \input{figures/eikgen_noperm_less1}
        \end{tikzpicture}
        \Bigg) \, .
\end{align}
The treatment of the case where $p_n$ in~\eqref{eq:eikid_gen_sum} corresponds to the loop momentum $\ell_m$ is independent of the number of external photons in the process. The discussion from Appendix A.2 of~\cite{Engel:2023ifn} therefore still applies and is summarised in the following. The loop momentum $\ell_m$ completely factorises in one of the diagrams on the r.h.s.\ of~\eqref{eq:eikid_gen_sum}, which renders the corresponding contribution scaleless and thus vanishing.\footnote{This holds since diagrams with more than one NLP vertex only contribute beyond NLP. Hence, wherever the open photon line for $\ell_m$ attaches, the on-shell eikonal identity~\eqref{eq:eikid_conv} applies and no scale dependence is introduced.} Which diagram vanishes depends on whether $\ell_m$ connects to a different external leg, to the external leg labelled by $v$, or to another open photon line to form a loop. If $\ell_m$ connects to a different external leg, it is the second diagram in~\eqref{eq:eikid_gen_sum} that is scaleless. This implies the simplified recursion relation
\begin{align} \label{eq:recursion1}
    \mathcal{T}_{n,m}^{(q)}
    = \frac{-Q v^{\mu_m}}{\ell_m \cdot v} \mathcal{T}_{n,m-1}^{(q+\ell_m)} \, .
\end{align}
If $\ell_m$ connects to the external leg labelled by $v$, there is an implicit dependence on $\ell_m$ in the residual momentum $q$. The substitution $q \to \tilde{q}-\ell_m$ makes this dependence explicit and shows that the first diagram in~\eqref{eq:eikid_gen_sum} is scaleless. Taking the loop momentum as outgoing, \textit{i.e.} $\ell_m \to - \ell_m$, this yields
\begin{align} \label{eq:recursion2}
    \mathcal{T}_{n,m}^{(q)}
    = \frac{Q v^{\mu_m}}{-\ell_m \cdot v+i0} \mathcal{T}_{n,m-1}^{(\tilde{q}+\ell_m)} \, .
\end{align}
The $+i0$ is given explicitly here to emphasize that the minus sign in the propagator cannot be factored out without changing the prescription. The case where $\ell_m$ connects to another loop momentum $\ell_i$ is slightly more involved. The corresponding recursion relation is stated here without reiterating the derivation from~\cite{Engel:2023ifn}. Assuming without loss of generality $\ell_j = -\ell_{m-1}$, the relation is given by
\begin{align} \label{eq:recursion3}
    \mathcal{T}_{n,m}^{(q)}
    = \frac{-Q v^{\mu_m}}{\ell_m \cdot v+i0}
    \frac{Q v^{\mu_{m-1}}}{-\ell_{m-1} \cdot v+i0}
    \mathcal{T}_{n,m-2}^{(q+\ell_m)} \, .
\end{align}
This concludes the discussion of the recursion relations for $p_n=\ell_m$, which remain unchanged compared to single-photon emission. The treatment of the case $p_n = k_n$, on the other hand, differs depending on whether there is one or multiple external photons in the process. If there is only one external soft scale, the second diagram on the r.h.s.\ of~\eqref{eq:eikid_gen_sum} is scaleless because the only scale fully factorises. If there are multiple external photons, this conclusion does not hold anymore and both diagrams are to be taken into account. The general recursion relation for external photons is
\begin{align} \label{eq:recursion4}
    \mathcal{T}_{n,m}^{(q)}
    = \frac{Q v^{\nu_n}}{-k_n \cdot v}
    \Big(\mathcal{T}_{n-1,m}^{(q)} - \mathcal{T}_{n-1,m}^{(q-k_n)} \Big) \, ,
\end{align}
where the photon momentum is taken to be outgoing as in~\eqref{eq:process_ng}.

The recursive application of the relations~\eqref{eq:recursion1}, \eqref{eq:recursion2}, \eqref{eq:recursion3}, and \eqref{eq:recursion4} yields the off-shell eikonal identity
\begin{align}
\label{eq:eikid_gen}
    	\mathcal{T}_{n,m}^{(q)}
     	= \Bigg( \prod_{i=1}^m \frac{-s_i Q v^{\mu_i}}{s_i \ell_i \cdot v+i0} \Bigg)
        \prod_{a=1}^n \frac{Q v^{\nu_a}}{-k_a \cdot v}
        \sum_{J \subset \{1,...,n\}}    
       	 \frac{i (-1)^{|J|}}{\widetilde{\sum}_j \ell_j \cdot v - \sum_{j \in J} k_j \cdot v + \tilde{q} \cdot v} \, .
\end{align}
This is the generalisation of the single-emission identity given in~(A.10) of~\cite{Engel:2023ifn} to multi-photon emission. All external photon momenta $k_a$ are taken to be incoming. The loop momenta $\ell_i$ are outgoing if they attach to the external leg labelled by $v$ and incoming if they connect to a different external leg. If two $\ell_i$ form a loop, one momentum is taken to be incoming and the other one outgoing. In this case the modified sum $\widetilde{\sum}$ takes into account the incoming momentum (\textit{c.f.}~\eqref{eq:recursion3}). Furthermore, $\tilde{q}$ is derived from $p$ by removing all dependence on the loop momenta $\ell_i$ (\textit{c.f.}~\eqref{eq:recursion2}). Finally note that $s_i$ is $+1$ if $\ell_i$ is incoming and $-1$ otherwise. The causal $+i0$ prescription is explicitly displayed in~\eqref{eq:eikid_gen} to indicate that these sign factors do not cancel. 

\subsection{Single-photon emission} \label{sec:hlet_1photon}

This section considers the simple case of single-photon emission at tree level in HLET. The corresponding amplitude is given by
\begin{align} \label{eq:hlet_1g_diags}
	\mathcal{A}_{m+1}^{(0)}
     & =
     \Big( \sum_i \mathcal{A}_{k_1,\mathcal{O}_\text{LP}}^{i,(0)}
     + \mathcal{A}_{k_1,\mathcal{O}_\text{NLP}}^{i,(0)} \Big)
     + \mathcal{A}_{;k_1,\mathcal{O}_\text{NLP}}^{(0)}
     + \Big( \sum_i \mathcal{A}_{k_1,\text{mag}}^{i,(0)}
     + \mathcal{A}_{k_1,0\gamma}^{i,(0)}
     + \mathcal{A}_{k_1,1\gamma}^{i,(0)} \Big)
     + \mathcal{O}(\lambda)
	\nonumber \\ 
    & =  \Bigg(\,  \sum_i 
            \begin{tikzpicture}[scale=.8,baseline={(0,-.1)}]
        	   \input{figures/hlet_1gLP}
       	   \end{tikzpicture}
    + \begin{tikzpicture}[scale=.8,baseline={(0,-.1)}]
        	   \input{figures/hlet_1gNLP_ext}
       	   \end{tikzpicture}
	  \, \Bigg)
	   +
	   \begin{tikzpicture}[scale=.8,baseline={(0,-.1)}]
        	   \input{figures/hlet_1gNLP_int}
       	   \end{tikzpicture}
    \nonumber \\ & \quad
        + \Bigg(\, \sum_i
	   \begin{tikzpicture}[scale=.8,baseline={(0,-.1)}]
        	   \input{figures/hlet_1gNLP_mag}
       	   \end{tikzpicture}
        + \begin{tikzpicture}[scale=.8,baseline={(0,-.1)}]
        	   \input{figures/hlet_1gNLP_kin0g}
       	   \end{tikzpicture}
        + \begin{tikzpicture}[scale=.8,baseline={(0,-.1)}]
        	   \input{figures/hlet_1gNLP_kin1g}
       	   \end{tikzpicture}
        \, \Bigg)
        + \mathcal{O}(\lambda) \, .
\end{align}
The amplitudes on the r.h.s.\ of the first line are in one-to-one correspondence with the diagrams in the second and third line. The diagrams where the kinetic vertex with no photon ($0\gamma$) corrects an on-shell leg vanish and are not explicitly displayed. This follows directly from the Feynman rule~\eqref{eq:vertex_kin} with $q_j=0$. Furthermore, in the case of single emission at tree level there is no contribution from the third kinetic vertex in~\eqref{eq:vertex_kin} with two photons. Section~\ref{sec:eft} provides all ingredients to compute the diagrams in~\eqref{eq:hlet_1g_diags} with the exception of $\mathcal{A}_{;k_1,\mathcal{O}_\mathrm{NLP}}^{(0)}$. This remaining contribution can be determined via matching, \textit{i.e.}\ by equating the HLET amplitude with the QED result of Section~\ref{sec:qed_1photon}.

Based on the eikonal vertex rule in~\eqref{eq:eikonal_rules}, the first two amplitudes evaluate to
\begin{align} 
    \label{eq:hlet_1g_OLP_tree}
    \mathcal{A}_{k_1,\mathcal{O}_\text{LP}}^{i,(0)}
    &= \eik_i(k_1) \mathcal{A}_{\mathcal{O}_\text{LP}}^{(0)}(\{-k_1\})
    = \eik_i(k_1) \mathcal{A}_m^{(0)} \, ,
    \\
    \label{eq:hlet_1g_ONLPext_tree}
    \mathcal{A}_{k_1,\mathcal{O}_\text{NLP}}^{i,(0)}
    &= \eik_i(k_1) \mathcal{A}_{\mathcal{O}_\text{NLP}}^{(0)}(\{-k_1\})
    = \eik_i(k_1) \Big(-k_1 \cdot \frac{\partial}{\partial p_i} \Gamma_i^{(0)}\Big) u(p_i ) \,  .
\end{align}
The HLET amplitudes $\mathcal{A}_{\mathcal{O}_\text{LP}}^{(0)}$ and $\mathcal{A}_{\mathcal{O}_\text{NLP}}^{(0)}$ are expressed in terms of the QED (sub)amplitudes $ \mathcal{A}_m^{(0)}=\Gamma_i^{(0)} u(p_i)$ with the matching relations~\eqref{eq:nonrad_matching} and \eqref{eq:nonrad_matching_photon}. In the following, these relations are often used without explicit reference.

The Feynman rule for the magnetic vertex~\eqref{eq:vertex_mag} and the eikonal propagator in~\eqref{eq:eikonal_rules} yield
\begin{align} \label{eq:hlet_1g_mag_tree}
    \mathcal{A}_{k_1,\text{mag}}^{i,(0)}
    = \frac{Q_i C_\text{mag}}{4m_i} \frac{i}{-k_1 \cdot v_i} \Gamma_i^{(0)} (1+\slashed{v_i})  \sigma_{\mu \nu} u_{v_i} \epsilon_1^{\mu} (-k_1^{\nu}) \, .
\end{align}
Upon using the Dirac algebra and the Dirac equation $\slashed{v_i} u_{v_i} = u_{v_i}$, this can be rewritten in terms of the tensor
\begin{align} \label{eq:Htensor}
    H_1^\mu 
    = \gamma^\mu 
    - \frac{1}{k_1 \cdot v_i} \slashed{k_1} v_i^\mu
    - \frac{1}{k_1 \cdot v_i} \gamma^\mu \slashed{k_1}
\end{align}
as
\begin{align} \label{eq:hlet_1g_mag_tree_tensor}
    \mathcal{A}_{k_1,\text{mag}}^{i,(0)}
    = \frac{Q_i C_\text{mag}}{2m_i} \Gamma_i^{(0)} 
    \epsilon_1 \cdot H_1 u_{v_i} \, .
\end{align}
Due to the particular structure of $H^\mu_1$, the magnetic contribution~\eqref{eq:hlet_1g_mag_tree_tensor} vanishes at the level of the unpolarised squared amplitude, \textit{i.e.}\ after interfering with the LP eikonal term and summing over the spin of the fermion with momentum $p_i$. This follows from basic Dirac algebra as shown in Section~5.2.1 of~\cite{Engel:2022kde}. 

The Feynman rules for the kinetic vertices~\eqref{eq:vertex_kin} give
\begin{align}
    \label{eq:single_kin0g_tree}
    \mathcal{A}_{k_1,0\gamma}^{(0)} 
    &= \frac{1}{2 m_i} \eik_i(k_1) \frac{k_{1,\perp_i}^2}{k_1 \cdot v_1} \mathcal{A}_m^{(0)}
    = \frac{Q_i}{2 m_i} \epsilon_1 \cdot v_i \mathcal{A}_m^{(0)} \, ,
    \\
    \label{eq:single_kin1g_tree}
    \mathcal{A}_{k_1,1\gamma}^{(0)}
    &= \frac{Q_i}{2m_i} 
    \frac{\epsilon_1 \cdot k_{1,\perp_i}}{k_1 \cdot v_i} \mathcal{A}_m^{(0)}
    = - \frac{Q_i}{2 m_i} \epsilon_1 \cdot v_i \mathcal{A}_m^{(0)}\, ,
\end{align}
where the $\perp_i$-components are expanded according to~\eqref{eq:perpcomp}.
%
%
The total kinetic contribution therefore vanishes
\begin{align} \label{eq:hlet_1g_kin_total}
    \mathcal{A}_{k_1,\text{kin}}^{(0)}
    =  \mathcal{A}_{k_1,0\gamma}^{(0)}
    + \mathcal{A}_{k_1,1\gamma}^{(0)}
    = 0 \, .
\end{align}

The remaining amplitude $\mathcal{A}_{;k_1,\text{NLP}}^{(0)}$ can now be determined by matching to the QED result given by the sum of~\eqref{eq:singleNLP} and \eqref{eq:singleNLP_internal}. With $p_i = m_i v_i + \mathcal{O}(\lambda)$, this yields
\begin{align} \label{eq:matching_internal}
    \mathcal{A}_{;k_1,\text{NLP}}^{(0)}
    = \sum_i \Bigg( -Q_i \Big( \epsilon_1 \cdot \frac{\partial}{\partial p_i} \Gamma_i^{(0)} \Big) u(p_i)
    + \frac{Q_i \Gamma_i^{(0)} \slashed{k_1} \slashed{\epsilon_1} u(p_i)}{ 2k_1 \cdot p_i}
    - \mathcal{A}_{k_1,\text{mag}}^{i,(0)} \Bigg)
    + \mathcal{O}(\lambda)\, .
\end{align}
By construction, the tree-level LBK theorem~\eqref{eq:lbk_tree} is reproduced after adding~\eqref{eq:hlet_1g_OLP_tree}, \eqref{eq:hlet_1g_ONLPext_tree}, and \eqref{eq:matching_internal}, squaring the result and summing over spins and polarisations. This implies the HLET-QED relation
\begin{align} \label{eq:relation_hlet_qed_tree}
    \sum_\text{pol} \Big|
    \Big( \sum_i \mathcal{A}_{k_1,\mathcal{O}_\text{LP}}^{i,(0)}
    + \mathcal{A}_{k_1,\mathcal{O}_\text{NLP}}^{i,(0)} \Big)
    + \mathcal{A}_{;k_1,\mathcal{O}_\text{NLP}}^{(0)} \Big|^2
    + \mathcal{O}(\lambda^0)
    = (E(k_1) + D(k_1)) \mathcal{M}_m^{(0)} + \mathcal{O}(\lambda^0) \, ,
\end{align}
with $E$ and $D$ defined in \eqref{eq:eikonal_squared} and \eqref{eq:lbkop_squared}, respectively. The magnetic contribution~\eqref{eq:hlet_1g_mag_tree_tensor} as well as the kinetic terms~\eqref{eq:single_kin0g_tree} and \eqref{eq:single_kin1g_tree} cancel and do not contribute to~\eqref{eq:relation_hlet_qed_tree}.

\subsection{Double-photon emission} \label{sec:hlet_2photon}

Based on the previous section, it is now possible to reproduce the tree-level double-emission result of Section~\ref{sec:qed_2photon} by reducing as many terms as possible to single-emission objects. The double-emission amplitude in HLET is given by
\begin{align}
    \label{eq:hlet_2g_tree_diags}
	\mathcal{A}_{m+2}^{(0)}
    &= \Big( \sum_{i,j}  \mathcal{A}_{k_1 k_2,\mathcal{O}_\text{LP}}^{ij,(0)} +\mathcal{A}_{k_1 k_2,\mathcal{O}_\text{NLP}}^{ij,(0)} \Big) +  \Big( \sum_i \mathcal{A}_{k_1;k_2,\mathcal{O}_\text{NLP}}^{i,(0)} + (k_1 \leftrightarrow k_2) \Big)
    \nonumber \\ & \quad
    + \Big( \sum_{i,j} \mathcal{A}_{k_1 k_2,\text{mag}}^{ij,(0)}
    + \mathcal{A}_{k_1 k_2,0\gamma}^{ij,(0)}
    + \mathcal{A}_{k_1 k_2,1\gamma}^{ij,(0)}\Big)
    + \Big( \sum_i \mathcal{A}_{k_1 k_2,2\gamma}^{ii,(0)} \Big) + \mathcal{O}(\lambda^0))
    \nonumber \\
	&=  \Bigg(\,  \sum_{i,j} 
            \begin{tikzpicture}[scale=.8,baseline={(0,-.1)}]
        	   \input{figures/hlet_2gLP}
       	   \end{tikzpicture}
    + \begin{tikzpicture}[scale=.8,baseline={(0,-.1)}]
        	   \input{figures/hlet_2gNLP_ext}
       	   \end{tikzpicture}
	  \, \Bigg)
	   + \Bigg(\, \sum_i
	   \begin{tikzpicture}[scale=.8,baseline={(0,-.1)}]
        	   \input{figures/hlet_2gNLP_int}
       	   \end{tikzpicture}
         + (k_1 \leftrightarrow k_2) \, \Bigg)
    \nonumber \\ & \quad
        + \Bigg(\, \sum_{i,j}
	   \begin{tikzpicture}[scale=.8,baseline={(0,-.1)}]
        	   \input{figures/hlet_2gNLP_mag}
       	   \end{tikzpicture}
        + \begin{tikzpicture}[scale=.8,baseline={(0,-.1)}]
        	   \input{figures/hlet_2gNLP_kin0g}
       	   \end{tikzpicture}
        + \begin{tikzpicture}[scale=.8,baseline={(0,-.1)}]
        	   \input{figures/hlet_2gNLP_kin1g}
       	   \end{tikzpicture}
        + (k_1 \leftrightarrow k_2)
        \, \Bigg)
    \nonumber \\ & \quad
    + \Bigg( \, \sum_i
    \begin{tikzpicture}[scale=.8,baseline={(0,-.1)}]
        	   \input{figures/hlet_2gNLP_kin2g}
       	   \end{tikzpicture}
    \, \Bigg)
        + \mathcal{O}(\lambda^0) \, .
\end{align}
In the third diagram, the additional external photon with momentum $k_1$ only has to be taken into account at LP. This follows from internal emission not contributing at LP. As a consequence, the diagram reduces to the single-emission terms
\begin{align} \label{eq:hlet_2gNLPint_sameleg}
    \mathcal{A}_{k_1;k_2,\mathcal{O}_\text{NLP}}^{i,(0)}
    &= \eik_i(k_1) \mathcal{A}_{;k_2,\mathcal{O}_\text{NLP}}^{(0)} \, .
\end{align}
This is also the case for contributions where the two photons are emitted from different legs, \textit{i.e.}\ if $i \neq j$. The LP term reduces to
\begin{align}
    \label{eq:hlet_2g_LP}
    \mathcal{A}_{k_1 k_2,\mathcal{O}_\text{LP}}^{i \neq j,(0)}
    &= \eik_i(k_1) \eik_j(k_2) \mathcal{A}_m^{(0)}
\end{align}
and the NLP to
\begin{align}
    \mathcal{A}_{k_1 k_2,V}^{i \neq j,(0)}
    &= \eik_i(k_1) \mathcal{A}_{k_2,V}^{j,(0)}
    + \eik_j(k_2) \mathcal{A}_{k_1,V}^{i,(0)} \, ,
\end{align}
with $V\in\{\mathcal{O}_\text{NLP},\text{mag},0\gamma,1\gamma\}$. Note that for $V=\mathcal{O}_\text{NLP}$, the two emissions factorise due to the linear dependence of the matching relation~\eqref{eq:nonrad_matching_photon} on the residual momenta $q_j$.

As for QED, new structures can only arise if the two photons are emitted from the same leg. This does not happen, however, in the case of the first two amplitudes in~\eqref{eq:hlet_2g_tree_diags} due to the on-shell eikonal identity~\eqref{eq:eikid_conv}, which yields
\begin{align}
    \label{eq:hlet_2gLP_sameleg}
    \mathcal{A}_{k_1 k_2,\mathcal{O}_\text{LP}}^{i i,(0)}
    &= \begin{tikzpicture}[scale=.8,baseline={(0,-.1)}]
        	   \input{figures/hlet_2gLP_sameleg}
       	   \end{tikzpicture}
    = \eik_i(k_1) \eik_i(k_2) \mathcal{A}_m^{(0)}
\end{align}
and
\begin{align}
    \label{eq:hlet_2gNLPext_sameleg}
    \mathcal{A}_{k_1 k_2,\mathcal{O}_\text{NLP}}^{i i,(0)}
    &= \begin{tikzpicture}[scale=.8,baseline={(0,-.1)}]
        	   \input{figures/hlet_2gNLP_sameleg}
       	   \end{tikzpicture}
    = \eik_i(k_1) \mathcal{A}_{k_2,\mathcal{O}_\text{NLP}}^{i,(0)}
    + \eik_i(k_2) \mathcal{A}_{k_1,\mathcal{O}_\text{NLP}}^{i,(0)} \, .
\end{align}

Furthermore, also the magnetic vertex does not give rise to a new structure and instead reduces to the single-emission terms
\begin{align}
    \mathcal{A}_{k_1 k_2,\text{mag}}^{ii,(0)}
    &= \begin{tikzpicture}[scale=.8,baseline={(0,-.1)}]
        	   \input{figures/hlet_magii1}
       	   \end{tikzpicture}
    + \begin{tikzpicture}[scale=.8,baseline={(0,-.1)}]
        	   \input{figures/hlet_magii2}
       	   \end{tikzpicture}
    + (k_1 \leftrightarrow k_2)
    \nonumber \\
    &= \eik_i(k_2) \mathcal{A}_{k_1,\text{mag}}^{i,(0)}
    + \eik_i(k_1) \mathcal{A}_{k_2,\text{mag}}^{i,(0)} \, ,
\end{align}
with $\mathcal{A}_{k_a,\text{mag}}^{i,(0)}$ given in~\eqref{eq:hlet_1g_mag_tree}.
This follows after rewriting the propagators in the second diagram as
\begin{align} \label{eq:partial_fraction_new}
    \frac{1}{k_1 \cdot v_i(k_1 \cdot v_i + k_2 \cdot v_i)}
    = \frac{1}{k_1 \cdot v_i k_2 \cdot v_i}
    - \frac{1}{k_2 \cdot v_i(k_1 \cdot v_i + k_2 \cdot v_i)} \, .
\end{align}

The only non-trivial contributions come from the kinetic vertices. The $0\gamma$-vertex gives rise to the diagrams
\begin{align} \label{eq:hlet_2g_kin0g_diag}
    \mathcal{A}_{k_1 k_2,0\gamma}^{ii,(0)}
    &= \begin{tikzpicture}[scale=.8,baseline={(0,-.1)}]
        	   \input{figures/hlet_kin0g_ii_1}
       	   \end{tikzpicture}
    + \Bigg( \, \begin{tikzpicture}[scale=.8,baseline={(0,-.1)}]
        	   \input{figures/hlet_kin0g_ii_2}
       	   \end{tikzpicture}
        + (k_1 \leftrightarrow k_2) \, \Bigg) \, .
\end{align}
After applying the on-shell eikonal identity~\eqref{eq:eikid_conv} to the first diagram and~\eqref{eq:partial_fraction_new} to the second one, the amplitude factorises according to
\begin{align} \label{eq:hlet_2g_kin0g}
    \mathcal{A}_{k_1 k_2,0\gamma}^{ii,(0)}
    &= \frac{1}{2 m_i} \eik_i(k_1) \eik_i(k_2) \mathcal{A}_m^{(0)}
     \Bigg( \frac{k_{1,\perp_i}^2}{k_1 \cdot v_i}
    + \frac{k_{2,\perp_i}^2}{k_2 \cdot v_i}
    + \frac{(k_1+k_2)_{\perp_i}^2-k_{1,\perp_i}^2-k_{2,\perp_i}^2}{k_1 \cdot v_i + k_2 \cdot v_i}\Bigg)
    \nonumber \\
    &= \eik_i(k_2) \mathcal{A}_{k_1,0\gamma}^{i,(0)} 
    + \eik_i(k_1) \mathcal{A}_{k_2,0\gamma}^{i,(0)}
    + \mathcal{G}_{i,0\gamma}^{(0)}(k_1,k_2) \mathcal{A}_m^{(0)}
\end{align}
into the single-emission amplitude $\mathcal{A}_{k_a,0\gamma}^{i,(0)} $ given in~\eqref{eq:single_kin0g_tree} and the genuine two-photon term
\begin{align} \label{eq:hlet_gikin0}
    \mathcal{G}_{i,0\gamma}^{(0)}(k_1,k_2)
    = \frac{Q_i^2}{m_i} \frac{\epsilon_1 \cdot v_i \epsilon_2 \cdot v_i k_{1,\perp_i} \cdot k_{2,\perp_i}}{k_1 \cdot v_i k_2 \cdot v_i (k_1 \cdot v_i + k_2 \cdot v_i)} \, .
\end{align}
Similarly, the $1\gamma$-vertex amplitude evaluates to
\begin{align}
    \mathcal{A}_{k_1 k_2,1\gamma}^{ii,(0)}
    &= \begin{tikzpicture}[scale=.8,baseline={(0,-.1)}]
        	   \input{figures/hlet_kin1g_ii_1}
       	   \end{tikzpicture}
    + \begin{tikzpicture}[scale=.8,baseline={(0,-.1)}]
        	   \input{figures/hlet_kin1g_ii_2}
       	   \end{tikzpicture}
    + (k_1 \leftrightarrow k_2)
    \nonumber \\ \label{eq:hlet_2g_kin1g}
    &= \eik_i(k_2) \mathcal{A}_{k_1,1\gamma}^{i,(0)}
    + \eik_i(k_1) \mathcal{A}_{k_2,1\gamma}^{i,(0)}
    +  \mathcal{G}_{i,1\gamma}^{(0)}(k_1,k_2)
    \mathcal{A}_m^{(0)} \, ,
\end{align}
with $\mathcal{A}_{k_a,1\gamma}^{i,(0)}$ given in~\eqref{eq:single_kin1g_tree} and
\begin{align} \label{eq:hlet_gikin1}
    \mathcal{G}_{i,1\gamma}^{(0)}(k_1,k_2) = - \frac{Q_i^2}{m_i} \frac{\epsilon_1 \cdot v_i \epsilon_2 \cdot k_{1,\perp_i} k_2 \cdot v_i + \epsilon_2 \cdot v_i \epsilon_1 \cdot k_{2,\perp_i} k_1 \cdot v_i}{k_1 \cdot v_i k_2 \cdot v_i (k_1 \cdot v_i + k_2 \cdot v_i)} \, .
\end{align}
The $2\gamma$-vertex does not contribute to single emission (at tree level) and therefore only generates the genuine double-emission contribution
\begin{align} \label{eq:hlet_2g_kin2g}
    \mathcal{A}_{k_1 k_2,2\gamma}^{ii,(0)}
    &= \begin{tikzpicture}[scale=.8,baseline={(0,-.1)}]
        	   \input{figures/hlet_2gNLP_kin2g}
       	   \end{tikzpicture}
    = \frac{Q_i^2}{m_i} \frac{\epsilon_{1,\perp_i} \cdot \epsilon_{2,\perp_i} k_1 \cdot v_i k_2 \cdot v_i}
    {k_1 \cdot v_i k_2 \cdot v_i (k_1 \cdot v_i + k_2 \cdot v_i)} \mathcal{A}_m^{(0)}
    = \mathcal{G}_{i,2\gamma}^{(0)}(k_1,k_2) \mathcal{A}_m^{(0)}\, .
\end{align}
Summing up the three kinetic contributions~\eqref{eq:hlet_2g_kin0g}, \eqref{eq:hlet_2g_kin1g}, and \eqref{eq:hlet_2g_kin2g}, writing out the $\perp_i$ components, and replacing $v_i = p_i/m_i + \mathcal{O}(\lambda)$ yields
\begin{align} \label{eq:amp_kin_tree_double}
    \mathcal{A}_{k_1 k_2,\text{kin}}^{ii,(0)}
    = \mathcal{A}_{k_1 k_2,0\gamma}^{ii,(0)}
    + \mathcal{A}_{k_1 k_2,1\gamma}^{ii,(0)}
    + \mathcal{A}_{k_1 k_2,2\gamma}^{ii,(0)}
    =\mathcal{G}_i^{(0)}(k_1,k_2) \mathcal{A}_m^{(0)} \, ,
\end{align}
where $\mathcal{G}_i^{(0)} = \mathcal{G}_{i,0\gamma}^{(0)} + \mathcal{G}_{i,1\gamma}^{(0)} +  \mathcal{G}_{i,2\gamma}^{(0)}$ is defined in~\eqref{eq:eiknts_amp} and the single-emission terms cancel each other according to~\eqref{eq:hlet_1g_kin_total}. Combining~\eqref{eq:amp_kin_tree_double} with the expressions for the $\mathcal{O}_\text{LP}$ and $\mathcal{O}_\text{NLP}$ amplitudes given in~\eqref{eq:hlet_2gNLPint_sameleg}, \eqref{eq:hlet_2gLP_sameleg}, and \eqref{eq:hlet_2gNLPext_sameleg} yields for the total NLP amplitude
\begin{align} \label{eq:hlet_lbk_2g_tree_amp}
    \mathcal{A}_{m+2}^{(0)}
    &= \eik(k_1) \eik(k_2) \mathcal{A}_m^{(0)}
    \nonumber \\ & \quad
    +  \Bigg( \eik(k_2) \Big(
         \sum_i \big( \mathcal{A}_{k_1,\mathcal{O}_\text{NLP}}^{i,(0)} 
        + \mathcal{A}_{k_1,\text{mag}}^{i,(0)} \big)
        + \mathcal{A}_{;k_1,\mathcal{O}_\text{NLP}}^{(0)}
        \Big)
    + (k_1 \leftrightarrow k_2) \Bigg)
    \nonumber \\ & \quad
    +  \mathcal{G}^{(0)}(k_1,k_2) \mathcal{A}_m^{(0)}
    + \mathcal{O}(\lambda^0) \, ,
\end{align}
with $\eik$ and $\mathcal{G}^{(0)}$ defined in~\eqref{eq:eik_sum}. This exactly corresponds to the QED result~\eqref{eq:lbk_2g_tree_amp} since the NLP single-emission contribution~\eqref{eq:single_nlp} in HLET is given by
\begin{align} \label{eq:single_nlp_hlet}
     \mathcal{A}_{k_a,\text{NLP}}^{(0)} 
     = \sum_i \big( \mathcal{A}_{k_a,\mathcal{O}_\text{NLP}}^{i,(0)} 
        + \mathcal{A}_{k_a,\text{mag}}^{i,(0)} \big)
        + \mathcal{A}_{;k_a,\mathcal{O}_\text{NLP}}^{(0)} \, .
\end{align}
Hence, the unpolarised squared amplitude~\eqref{eq:lbk_2g_tree} is successfully reproduced in HLET. 

\subsection{Triple-photon emission} \label{sec:hlet_3photon}

Before turning to the case of an arbitrary number of external photons, the triple-emission result from Section~\ref{sec:qed_3photon} is reproduced in the following. Again the strategy is to reduce as many terms as possible to lower multiplicity contributions. The NLP triple-emission amplitude is given by
\begin{align} \label{eq:hlet_3g_tree_diags}
	\mathcal{A}_{m+3}^{(0)}
    &= \Big( \sum_{i,j,l}  \mathcal{A}_{k_1 k_2 k_3,\mathcal{O}_\text{LP}}^{ijl,(0)} +\mathcal{A}_{k_1 k_2 k_3,\mathcal{O}_\text{NLP}}^{ijl,(0)} \Big) 
    +  \Big( \sum_i \mathcal{A}_{k_1 k_2;k_3,\mathcal{O}_\text{NLP}}^{ij,(0)} 
            + (k_{1,2} \leftrightarrow k_3) \Big)
    \nonumber \\ & \quad
    + \Big( \sum_{i,j,l} \mathcal{A}_{k_1 k_2 k_3,\text{mag}}^{ijl,(0)}
    + \mathcal{A}_{k_1 k_2 k_3,0\gamma}^{ijl,(0)}
    + \mathcal{A}_{k_1 k_2 k_3,1\gamma}^{ijl,(0)}\Big)
    + \Big( \sum_{i,l} \mathcal{A}_{k_1 k_2; k_3,2\gamma}^{iil,(0)} 
            + (k_{1,2} \leftrightarrow k_3)\Big)
    + \mathcal{O}(\lambda^0)) \, ,
\end{align}
with the diagrammatic representation
\begin{align}
	\mathcal{A}_{m+3}^{(0)}
	& =  \Bigg(\,  \sum_{i,j,l} 
            \begin{tikzpicture}[scale=.8,baseline={(0,-.1)}]
        	   \input{figures/hlet_3gLP}
       	   \end{tikzpicture}
    + \begin{tikzpicture}[scale=.8,baseline={(0,-.1)}]
        	   \input{figures/hlet_3gNLP_ext}
       	   \end{tikzpicture}
	  \, \Bigg)
	   + \Bigg(\, \sum_{i,j}
	   \begin{tikzpicture}[scale=.8,baseline={(0,-.1)}]
        	   \input{figures/hlet_3gNLP_int}
       	   \end{tikzpicture}
         + (k_{1,2} \leftrightarrow k_3) \, \Bigg)
    \nonumber \\ & \quad
        + \Bigg(\, \sum_{i,j,l}
	   \begin{tikzpicture}[scale=.8,baseline={(0,-.1)}]
        	   \input{figures/hlet_3gNLP_mag}
       	   \end{tikzpicture}
        + \begin{tikzpicture}[scale=.8,baseline={(0,-.1)}]
        	   \input{figures/hlet_3gNLP_kin0g}
       	   \end{tikzpicture}
        + \begin{tikzpicture}[scale=.8,baseline={(0,-.1)}]
        	   \input{figures/hlet_3gNLP_kin1g}
       	   \end{tikzpicture}
        + (k_{2,3} \leftrightarrow k_1)
        \, \Bigg)
    \nonumber \\ & \quad
    + \Bigg( \, \sum_{i,l}
    \begin{tikzpicture}[scale=.8,baseline={(0,-.1)}]
        	   \input{figures/hlet_3gNLP_kin2g}
       	   \end{tikzpicture}
    + (k_{1,2} \leftrightarrow k_3)
    \, \Bigg)
        + \mathcal{O}(\lambda^0) \, .
\end{align}
The shorthand notation $(k_{i,j} \leftrightarrow k_l)$ is defined in~\eqref{eq:2perm_short}. The same logic as in the double-emission case of the previous section applies here. The diagrams in the first line trivially reduce to eikonal factors times single-emission terms as an immediate consequence of the on-shell eikonal identity~\eqref{eq:eikid_conv}. All remaining diagrams reduce to double- and single-emission contributions if the three photons are not emitted from the same leg. As a consequence, only diagrams with triple emission from the same leg have to be considered explicitly.

The magnetic amplitude receives contributions from the diagrams
\begin{align} \label{hlet_3g_mag_res_diag}
    \mathcal{A}_{k_1 k_2 k_3,\text{mag}}^{iii,(0)}
    &= \begin{tikzpicture}[scale=.8,baseline={(0,-.1)}]
        	   \input{figures/hlet_3gNLP_mag_sameleg_1}
       	   \end{tikzpicture}
    +  \begin{tikzpicture}[scale=.8,baseline={(0,-.1)}]
        	   \input{figures/hlet_3gNLP_mag_sameleg_2}
       	   \end{tikzpicture} 
    + \Bigg( \, 
        \begin{tikzpicture}[scale=.8,baseline={(0,-.1)}]
        	   \input{figures/hlet_3gNLP_mag_sameleg_3}
       	   \end{tikzpicture}
        + (k_1 \leftrightarrow k_2)\, \Bigg)
    \nonumber \\ & \quad
    + (k_{1,2} \leftrightarrow k_3) \, .
\end{align}
Applying the on-shell eikonal identity~\eqref{eq:eikid_conv} to the first diagram and the off-shell eikonal identity~\eqref{eq:eikid_gen} to the latter two, they evaluate to
\begin{align} \label{hlet_3g_mag_res}
    \mathcal{A}_{k_1 k_2 k_3,\text{mag}}^{iii,(0)}
    &= k_3 \cdot v_i \mathcal{A}_{k_3,\text{mag}}^{i,(0)}
    \eik_i(k_1) \eik_i(k_2)
    \Big(
        \frac{1+1-(1+1)}{k_1 \cdot v_i + k_2 \cdot v_i + k_3 \cdot v_i}
        + \frac{0-1+(1+0)}{k_1 \cdot v_i + k_3 \cdot v_i}
    \nonumber \\ &
        + \frac{0-1+(0+1)}{k_2 \cdot v_i + k_3 \cdot v_i}
        + \frac{0+1+(0+0)}{k_3 \cdot v_i}
    \Big)
    + (k_{1,2} \leftrightarrow k_3)
    \nonumber \\
    &= \eik_i(k_1) \eik_i(k_2) \mathcal{A}_{k_3,\text{mag}}^{i,(0)}
    + (k_{1,2} \leftrightarrow k_3) \, .
\end{align}
The first two lines of~\eqref{hlet_3g_mag_res} are written such that the individual contributions from the off-shell eikonal identity~\eqref{eq:eikid_gen} due to different choices of $J \subset \{k_1,k_2,k_3\}$ can easily be identified. This can be illustrated with the second term in the bracket. The two zeros in the numerator indicate that the first and fourth (corresponding to the $(k_1 \leftrightarrow k_2)$ term) diagram in the first line of~\eqref{hlet_3g_mag_res_diag} do not contribute to this propagator structure. The second and third term in the numerator correspond to the contributions of the second and third diagram with $J=\{k_1\}$ and $J=\{\}$, respectively. As in the case of double emission, the magnetic contribution reduces to single-emission terms and thus vanishes for unpolarised scattering.

Similarly, the $0\gamma$-amplitude evaluates to
\begin{align}
    \mathcal{A}_{k_1 k_2 k_3,0\gamma}^{iii,(0)}
    &= \begin{tikzpicture}[scale=.8,baseline={(0,-.1)}]
        	   \input{figures/hlet_3gNLP_kin0g_sameleg_1}
       	   \end{tikzpicture}
    + \Bigg( \, 
        \begin{tikzpicture}[scale=.8,baseline={(0,-.1)}]
        	   \input{figures/hlet_3gNLP_kin0g_sameleg_2}
       	   \end{tikzpicture} 
    + \begin{tikzpicture}[scale=.8,baseline={(0,-.1)}]
        	   \input{figures/hlet_3gNLP_kin0g_sameleg_3}
       	   \end{tikzpicture}
        + (k_{1,2} \leftrightarrow k_3)\, \Bigg)
    \nonumber \\
    &= \frac{1}{2m_i}\eik_i(k_1) \eik_i(k_2) \eik_i(k_3) \mathcal{A}_m^{(0)}
    \Bigg(
        \frac{
        (k_1+k_2+k_3)_{\perp_i}^2
        - \big((k_1+k_2)_{\perp_i}^2 - k_{3,\perp_i}^2 + (k_{1,2} \leftrightarrow k_3) \big)
        }{k_1 \cdot v_i + k_2 \cdot v_i + k_3 \cdot v_i}
    \nonumber \\ &
        + \Big( \frac{
        (k_1+k_2)_{\perp_i}^2
        - k_{1,\perp_i}^2 - k_{2,\perp_i}^2
        }{k_1 \cdot v_i + k_2 \cdot v_i}
        + \frac{k_{3,\perp_i}}{k_3 \cdot v_i}
        + (k_{1,2} \leftrightarrow k_3)
        \Big)
    \Bigg) \, .
\end{align}
The numerator of the first fraction vanishes and the expression simplifies to
\begin{align} \label{eq:triple_kin0g_tree}
    \mathcal{A}_{k_1 k_2 k_3,0\gamma}^{iii,(0)}
    = \eik_i(k_1) \eik_i(k_2) \mathcal{A}_{k_3,0\gamma}^{i,(0)}
    + \eik_i(k_3) \mathcal{G}_{i,0\gamma}^{(0)}(k_1,k_2) \mathcal{A}_m^{(0)}
    + (k_{1,2} \leftrightarrow k_3) \, ,
\end{align}
with the single-emission amplitude $\mathcal{A}_{k_a,0\gamma}^{i,(0)}$ defined in~\eqref{eq:single_kin0g_tree} and the genuine double-emission contribution $\mathcal{G}_{i,0\gamma}^{(0)}$ given in~\eqref{eq:hlet_gikin0}.
Finally, the results for the two remaining kinetic amplitudes read
\begin{align}
    \mathcal{A}_{k_1 k_2 k_3,1\gamma}^{iii,(0)}
    &= \begin{tikzpicture}[scale=.8,baseline={(0,-.1)}]
        	   \input{figures/hlet_3gNLP_kin1g_sameleg_1}
       	   \end{tikzpicture}
    +  \begin{tikzpicture}[scale=.8,baseline={(0,-.1)}]
        	   \input{figures/hlet_3gNLP_kin1g_sameleg_2}
       	   \end{tikzpicture} 
    + \Bigg( \, 
        \begin{tikzpicture}[scale=.8,baseline={(0,-.1)}]
        	   \input{figures/hlet_3gNLP_kin1g_sameleg_3}
       	   \end{tikzpicture}
        + (k_1 \leftrightarrow k_2)\, \Bigg)
    \nonumber \\ & \quad
    + (k_{1,2} \leftrightarrow k_3)
    \nonumber \\
    & = \eik_i(k_1) \eik_i(k_2) \mathcal{A}_{k_3,1\gamma}^{i,(0)}
    + \eik_i(k_3) \mathcal{G}_{i,1\gamma}^{(0)}(k_1,k_2) \mathcal{A}_m^{(0)}
    + (k_{1,2} \leftrightarrow k_3) \, ,
    \\
    \mathcal{A}_{k_1 k_2 k_3,2\gamma}^{iii,(0)}
    &= \begin{tikzpicture}[scale=.8,baseline={(0,-.1)}]
        	   \input{figures/hlet_3gNLP_kin2g_sameleg_1}
       	   \end{tikzpicture}
    +  \begin{tikzpicture}[scale=.8,baseline={(0,-.1)}]
        	   \input{figures/hlet_3gNLP_kin2g_sameleg_2}
       	   \end{tikzpicture} 
    + (k_{1,2} \leftrightarrow k_3)
    \nonumber \\
    &= \eik_i(k_3) \mathcal{G}_{i,2\gamma}^{(0)}(k_1,k_2) \mathcal{A}_m^{(0)}
    + (k_{1,2} \leftrightarrow k_3) \, .
\end{align}
The single emission amplitude $\mathcal{A}_{k_a,1\gamma}^{i,(0)}$ is given in~\eqref{eq:single_kin0g_tree} and the two double-emission objects $\mathcal{G}_{i,1\gamma}^{(0)}(k_1,k_2)$ and $\mathcal{G}_{i,2\gamma}^{(0)}(k_1,k_2) $ are defined in~\eqref{eq:hlet_gikin1} and \eqref{eq:hlet_2g_kin2g}, respectively. In total, the kinetic contribution therefore reads
\begin{align}
    \mathcal{A}_{k_1 k_2 k_3,\text{kin}}^{iii,{(0)}}
    = \mathcal{A}_{k_1 k_2 k_3,0\gamma}^{iii,{(0)}}
    + \mathcal{A}_{k_1 k_2 k_3,1\gamma}^{iii,{(0)}}
    + \mathcal{A}_{k_1 k_2 k_3,2\gamma}^{iii,{(0)}}
    = \eik_i(k_3) \mathcal{G}_i^{(0)}(k_1,k_2) + (k_{1,2} \leftrightarrow k_3) \, ,
\end{align}
with $\mathcal{G}_i^{(0)} = \mathcal{G}_{i,0\gamma}^{(0)} + \mathcal{G}_{i,1\gamma}^{(0)} +  \mathcal{G}_{i,2\gamma}^{(0)}$ defined in~\eqref{eq:eiknts_amp}. As for double emission, the single-emission terms cancel according to~\eqref{eq:hlet_1g_kin_total}. 
All of these results combined exactly reproduce the QED amplitude~\eqref{eq:lbk_3g_tree_amp} and therefore also the soft theorem~\eqref{eq:lbk_ng_allorder} for triple emission at tree level.

\subsection{$n$-photon emission} \label{sec:hlet_nphoton}

This section formalises the previous HLET calculations and extends them to an arbitrary number of external photons. At LP the most general diagram to be considered is
\begin{align} \label{eq:hlet_ngLP_general_tree}
    \mathcal{A}_{k_1...k_n,\mathcal{O}_\text{LP}}^{i...i,(0)}
    &= \begin{tikzpicture}[scale=.8,baseline={(0,-.1)}]
        	   \input{figures/hlet_ngLP_general}
       	   \end{tikzpicture} \, ,
\end{align}
where the dashed line denotes $n$ photons attached to the crossed vertex. Emissions from different legs trivially factorise and are not considered explicitly here. Upon applying the on-shell eikonal identity~\eqref{eq:eikid_conv}, this amplitude takes the simple form
\begin{align}
    \mathcal{A}_{k_1...k_n,\mathcal{O}_\text{LP}}^{i...i,(0)}
    = \prod_{a=1}^{n} \eik_i(k_a) \mathcal{A}_m^{(0)} \, .
\end{align}
Exactly the same reasoning applies to the $\mathcal{O}_{\text{NLP}}$ diagrams
\begin{align} 
    \label{eq:hlet_ngNLPint_general_tree}
    \mathcal{A}_{k_1...k_{n-1};k_n,\mathcal{O}_\text{NLP}}^{i...i,(0)}
    &= \begin{tikzpicture}[scale=.8,baseline={(0,-.1)}]
        	   \input{figures/hlet_ngNLP_int_general}
       	   \end{tikzpicture}
    = \prod_{c=1}^{n-1} \eik_i(k_c) \mathcal{A}_{;k_n,\mathcal{O}_\text{NLP}}^{(0)} \, ,
    \\
    \label{eq:hlet_ngNLPext_general_tree}
    \mathcal{A}_{k_1...k_n,\mathcal{O}_\text{NLP}}^{i...i,(0)}
    &= \begin{tikzpicture}[scale=.8,baseline={(0,-.1)}]
        	   \input{figures/hlet_ngNLP_ext_general}
       	   \end{tikzpicture}
    = \sum_{a=1}^n\ \prod_{c \neq a} \eik_i(k_c) \mathcal{A}_{k_a,\mathcal{O}_\text{NLP}}^{i,(0)} \, .
\end{align}

The remaining NLP diagrams with insertions of the magnetic and the kinetic vertices are more involved. The corresponding generic diagram is given by
\begin{align} \label{eq:hlet_ngNLP_general_tree}
    \mathcal{A}_{p,V}^{n,i}
    &= \sum_{K \subset N} \begin{tikzpicture}[scale=.8,baseline={(0,-.1)}]
        	   \input{figures/hlet_ngNLP_general}
       	   \end{tikzpicture}
    + (k_{1,...,p} \leftrightarrow k_{p+1,...,n}) \, ,
\end{align}
where the sets $P = \{1,...,p\}$, $K \subset N = \{p+1,...,n\}$, and $\bar{K}=N \setminus K$ define the index sets for the momenta $k_a$ connected to the respective vertex. The notation $(k_{1,...,p} \leftrightarrow k_{p+1,...,n})$ is a generalisation of~\eqref{eq:2perm_short} and denotes all possible permutations of $\{k_1,...k_n\}$ modulo permutations of $\{k_1,...,k_p\}$ and $\{k_{p+1},...,k_n\}$. The magnetic and kinetic contributions are related to this generic amplitude via
\begin{align}
    \mathcal{A}_{k_1...k_n,\text{mag}}^{i...i,(0)} \label{eq:relation_gen2hlet_mag}
    &= \mathcal{A}_{1,\text{mag}}^{n,i} \, ,
    \\
    \mathcal{A}_{k_1...k_n,0\gamma}^{i...i,(0)}
    &= \mathcal{A}_{0,0\gamma}^{n,i} \, ,
    \\
    \mathcal{A}_{k_1...k_n,1\gamma}^{i...i,(0)}
    &= \mathcal{A}_{1,1\gamma}^{n,i} \, ,
    \\
    \mathcal{A}_{k_1...k_n,2\gamma}^{i...i,(0)}
    &= \mathcal{A}_{2,2\gamma}^{n,i} \, .
\end{align}

The on-shell eikonal identity~\eqref{eq:eikid_conv} can be used for the dashed line in~\eqref{eq:hlet_ngNLP_general_tree} with the index set $K$. In the case of the $\bar{K}$ line, the generalised off-shell identity~\eqref{eq:eikid_gen} applies with
\begin{align}
    \tilde{p} \cdot v_i = - \sum_{j \in K} k_j \cdot v_i - \sum_{j=1}^p k_j \cdot v_i
\end{align}
and $m=0$. This yields for the amplitude~\eqref{eq:hlet_ngNLP_general_tree} the expression
\begin{align} \label{eq:amp_NLP_generic_tree}
    \mathcal{A}_{p,V}^{n,i}
    = \Gamma_i^{(0)} \Big( \prod_{c=1}^p \epsilon_c^{\rho_{c}}
    \prod_{c=p+1}^n \eik_i(k_c)
    \sum_{K \subset N} F^{K,V}_{\rho_1...\rho_p} 
    \sum_{J \subset \bar{K}} (-1)^{|J|} \mathcal{P}^{\tilde{J}}_{p}
    \Big) u_{v_i} + (k_{1,...,p} \leftrightarrow k_{p+1,...,n}) \, ,
\end{align}
with $\tilde{J} = K \cup J$ and the propagator structure
\begin{align}
    \mathcal{P}^{\tilde{J}}_{p}
    = \frac{i}{-\sum_{j \in \tilde{J}} k_j \cdot v_i 
         - \sum_{j=1}^{p} k_j \cdot v_i} \, .
\end{align}
The Feynman rules for the magnetic and kinetic vertices given in~\eqref{eq:vertex_mag} and \eqref{eq:vertex_kin} imply for the corresponding vertex factors
\begin{align}
    F^{K,\text{mag}}_{\rho_1} \label{eq:vertexfac_tree_mag_interm}
    &= -\frac{Q_i C_{\text{mag}}}{4 m_i}(1+\slashed{v}_i) \sigma_{\rho_1 \lambda} k_1^\lambda \, ,
    \\
    F^{K,0\gamma} \label{eq:vertexfac_tree_kin0_interm}
    &= \frac{i}{2 m_i} \Big( -\sum_{j \in K} k_j \Big)_{\perp_i}^2 \, ,
    \\
    F^{K,1\gamma}_{\rho_1} \label{eq:vertexfac_tree_kin1_interm}
    &= \frac{-i Q_i}{2 m_i} \Big( -k_1 - 2 \sum_{j \in K} k_j \Big)^{\perp_i}_{\rho_1} \, ,
    \\
    F^{K,2\gamma}_{\rho_1 \rho_2} \label{eq:vertexfac_tree_kin2_interm}
    &= \frac{i Q_i^2}{m_i} g^{\perp_i}_{\rho_1 \rho_2} \, .
\end{align}
The expression~\eqref{eq:amp_NLP_generic_tree} contains different propagator structures labelled by $\tilde{J}$. In order to collect for these structures the two sums over $K$ and $J$ can be reshuffled according to
\begin{align} \label{eq:reshuffle_id}
    \sum_{K \subset N} a(K) 
    \sum_{J \subset \bar{K}} (-1)^{|J|}\, b(K \cup J)
    = \sum_{\tilde{J} \subset N} (-1)^{|\tilde{J}|}\, b(\tilde{J})
    \sum_{K \subset \tilde{J}} (-1)^K\, a(K) \, ,
\end{align}
where $a$ and $b$ are generic functions of the index sets. Identifying $a$ with the vertex factor $F_{\rho_1...\rho_p}^{K,V}$ in~\eqref{eq:amp_NLP_generic_tree} and $b$ with the propagator structure $\mathcal{P}_p^{\tilde{J}}$, the amplitude is rewritten as
\begin{align} \label{eq:amp_NLP_generic_tree_new}
    \mathcal{A}_{p,F}^{n,i}
    = \Gamma_i^{(0)} \Big( \prod_{c=1}^p \epsilon_c^{\rho_{c}} \prod_{c=p+1}^{n} \eik_i(k_c)
    \sum_{\tilde J \subset N} \bar{F}^{\tilde{J},V}_{\rho_1...\rho_{p}} \mathcal{P}_{p}^{\tilde{J}}
     \Big) u_{v_i} \, ,
\end{align}
where the coefficient of the propagator structure labelled by $\tilde{J}$ is given by
\begin{align} \label{eq:propcoeff_tree}
    \bar{F}^{\tilde{J},V}_{\rho_1...\rho_{p}}
    = (-1)^{|\tilde{J}|} \sum_{K \subset \tilde{J}} (-1)^{|K|} F^{K,V}_{\rho_1...\rho_{p}} \, .
\end{align}
It turns out that the sum in~\eqref{eq:propcoeff_tree} can be evaluated in closed form for any $\tilde{J}$ for all vertices. In particular, plugging~\eqref{eq:vertexfac_tree_mag_interm}-\eqref{eq:vertexfac_tree_kin2_interm} into~\eqref{eq:propcoeff_tree} gives
\begin{align}
    \bar{F}^{\tilde{J},\text{mag}}_{\rho_1} \label{eq:vertexfac_tree_mag}
    &= -\frac{Q_i C_{\text{mag}}}{4 m_i}(1+\slashed{v_i})\sigma_{\rho_1 \lambda} k_1^\lambda
    \begin{cases}
        1, & \tilde{J} = \{ \} \\
        0, & \text{otherwise}
    \end{cases} \, ,
    \\ 
    \bar{F}^{\tilde{J},0\gamma} \label{eq:vertexfac_tree_kin0g}
    &= \frac{i}{2 m_i} 
    \begin{cases}
        0, & \tilde{J} = \{ \} \\
        k_{a,\perp_i}^2, & \tilde{J} = \{k_a\} \\
        2 k_a^{\perp_i} \cdot k_b^{\perp_i}, & \tilde{J} = \{k_a,k_b\} \\
        0, & \text{otherwise}
    \end{cases} \, ,
    \\
    \bar{F}^{\tilde{J},1\gamma}_{\rho_1} \label{eq:vertexfac_tree_kin1g}
    &= \frac{i Q_i}{2 m_i} 
    \begin{cases}
        k_{1,\rho_1}^{\perp_i}, & \tilde{J} = \{ \} \\
        2 k_{a,\rho_1}^{\perp_i}, & \tilde{J} = \{k_a\} \\
        0, & \text{otherwise}
    \end{cases} \, ,
    \\
    \bar{F}^{\tilde{J},2\gamma}_{\rho_1\rho_2} \label{eq:vertexfac_tree_kin2g}
    &= \frac{i Q_i^2}{m_i} g^{\perp_i}_{\rho_1 \rho_2}
    \begin{cases}
        1, & \tilde{J} = \{ \} \\
        0, & \text{otherwise}
    \end{cases} \, .
\end{align}
Hence, these coefficients are only non-zero for very few choices of $\tilde{J}$. This results in an enormous simplification since the non-trivial part of the amplitude~\eqref{eq:amp_NLP_generic_tree_new} is parameterised by $\tilde{J}$ and $p$. The dependence on the photon momenta $k_a$ with $a \notin \tilde{J} \cup P$, on the other hand, trivially factorises in terms of eikonal factors. 

The amplitude~\eqref{eq:amp_NLP_generic_tree_new} can now be evaluated straightforwardly for the four vertex types. The magnetic coefficient~\eqref{eq:vertexfac_tree_mag} vanishes for $|\tilde{J}|>0$. Since in addition $p=1$ in~\eqref{eq:relation_gen2hlet_mag}, the amplitude~\eqref{eq:amp_NLP_generic_tree_new} reduces to single-emission terms times eikonal factors. In particular, it evaluates to
\begin{align} \label{eq:hlet_ng_tree_mag}
    \mathcal{A}_{k_1...k_n,\text{mag}}^{i...i,(0)}
    = \mathcal{A}_{1,\text{mag}}^{n,i}
    &= \Gamma_i^{(0)} \Big( \epsilon_1^{\rho_1} \prod_{c=2}^n \eik_i(k_c)
    \bar{F}^{\{\},\text{mag}}_{\rho_1} \mathcal{P}_1^{\{\}} \Big) u_{v_i}
    + (k_1 \leftrightarrow k_{2,...,n})
    \nonumber \\
    &= \prod_{c=2}^n \eik_i(k_c) \Big( \frac{Q_i C_\text{mag}}{4m_i} \frac{i}{k_1 \cdot v_i} \Gamma_i^{(0)} (1+\slashed{v_i})  \sigma_{\rho_1 \lambda} u_{v_i} \epsilon_1^{\rho_1} k_1^{\lambda} \Big)
    + (k_1 \leftrightarrow k_{2,...,n})
    \nonumber \\
    &= \sum_{a=1}^n \prod_{c \neq a} \eik_i(k_c) \mathcal{A}_{k_a,\text{mag}}^{i,(0)} \, ,
\end{align}
where the expression for the magnetic single-emission amplitude~\eqref{eq:hlet_1g_mag_tree} is recovered. This confirms the simple structure~\eqref{hlet_3g_mag_res} in the triple-emission case and extends it to an arbitrary number of photon emissions.

The kinetic coefficients~\eqref{eq:vertexfac_tree_kin0g}, \eqref{eq:vertexfac_tree_kin1g}, and \eqref{eq:vertexfac_tree_kin2g} all vanish for $p+|\tilde{J}|>2$. As a result, the corresponding amplitudes reduce to the single- and double-emission terms
\begin{align} \label{eq:hlet_ng_tree_magkin}
    \mathcal{A}_{k_1...k_n,V}^{i...i,(0)}
    &= \sum_{a=1}^n \prod_{c \neq a} \eik_i(k_c) \mathcal{A}_{k_a,V}^{i,(0)}
    + \sum_{a,b=1}^n \prod_{c \neq a,b} \eik_i(k_c)  \mathcal{G}_{i,V}^{(0)}(k_a,k_b) \mathcal{A}_m^{(0)} \, ,
\end{align}
with $V \in \{0\gamma,1\gamma,2\gamma\}$. The lower multiplicity objects $\mathcal{A}_{k_a,V}^{i,(0)}$ and $\mathcal{G}_{i,V}^{(0)}(k_a,k_b)$ have been calculated in Sections~\ref{sec:hlet_1photon} and~\ref{sec:hlet_2photon}, respectively. For the purpose of illustration, this is demonstrated explicitly in the case of the $0\gamma$-vertex. Plugging the coefficient~\eqref{eq:vertexfac_tree_kin0g} into the amplitude~\eqref{eq:amp_NLP_generic_tree_new} gives
\begin{align}
    \mathcal{A}_{k_1...k_n,0\gamma}^{i...i,(0)}
    &= \Gamma_i^{(0)} \prod_{c=1}^n \eik_i(k_c)
    \Big( \sum_{a=1}^n \bar{F}^{\{k_a\},0\gamma} \mathcal{P}_0^{\{k_a\}}
    + \sum_{a,b=1}^n \bar{F}^{\{k_a,k_b\},0\gamma} \mathcal{P}_0^{\{k_a,k_b\}} \Big) u_{v_i}
    \nonumber \\
    &= \prod_{c=1}^n \eik_i(k_c) \frac{1}{2 m_i}
    \Big( \sum_{a=1}^n \frac{k_{a,\perp_i}^2 }{k_a \cdot v_i}
    + \sum_{a,b=1}^n  \frac{2 k_{a,\perp_i} \cdot k_{b,\perp_i}}{k_a \cdot v_i + k_b \cdot v_i} \Big) \mathcal{A}_m^{(0)}
    \nonumber \\
    &= \sum_{a=1}^n \prod_{c \neq a} \eik_i(k_c) \mathcal{A}_{k_a,0\gamma}^{i,(0)}
    + \sum_{a,b=1}^n \prod_{c \neq a,b} \eik_i(k_c) \mathcal{G}_{i,0\gamma}^{(0)}(k_a,k_b) \mathcal{A}_m^{(0)} \, ,
\end{align}
recovering the single- and double-emission objects given in~\eqref{eq:single_kin0g_tree} and~\eqref{eq:hlet_gikin0}, respectively. 

The sum of the three kinetic amplitudes is thus given by
\begin{align}
    \mathcal{A}_{k_1...k_n,\text{kin}}^{i...i,(0)}
    = \mathcal{A}_{k_1...k_n,0\gamma}^{i...i,(0)}
    + \mathcal{A}_{k_1...k_n,1\gamma}^{i...i,(0)}
    + \mathcal{A}_{k_1...k_n,2\gamma}^{i...i,(0)}
    = \sum_{a,b=1}^n \prod_{c \neq a,b} \eik_i(k_c) \mathcal{G}_i^{(0)}(k_a,k_b) \mathcal{A}_m^{(0)} \, ,
\end{align}
with $\mathcal{G}_i^{(0)}$ defined in~\eqref{eq:eiknts_amp} and the single-emission term cancelling according to~\eqref{eq:hlet_1g_kin_total}. Together with the $\mathcal{O}_\text{LP}$ and $\mathcal{O}_\text{NLP}$ results in~\eqref{eq:hlet_ngLP_general_tree}, \eqref{eq:hlet_ngNLPint_general_tree}, and \eqref{eq:hlet_ngNLPext_general_tree} this yields for the $n$-emission amplitude
\begin{align}
    \mathcal{A}_{m+n}^{(0)}
    = \prod_{a=1}^n \eik(k_a) \mathcal{A}_m^{(0)}
    + \sum_{a=1}^n \prod_{c \neq a} \eik(k_c) \mathcal{A}_{k_a,\text{NLP}}^{(0)}
    + \sum_{a,b=1}^n \prod_{c \neq a,b} \eik(k_c) \mathcal{G}^{(0)}(k_a,k_b) \mathcal{A}_m^{(0)}
    + \mathcal{O}(\lambda^{-n+2}) \, ,
\end{align}
with $\eik$ and $\mathcal{G}^{(0)}$ defined in~\eqref{eq:eik_sum} and the total NLP single-emission contribution $\mathcal{A}_{k_a,\text{NLP}}^{(0)}$ given by~\eqref{eq:single_nlp_hlet}. This generalises the tree-level result~\eqref{eq:lbk_3g_tree_amp} for triple emission to an arbitrary number of soft photons and proves the soft theorem~\eqref{eq:lbk_ng_allorder} at tree level. Compared to QED, the simple NLP structure becomes much more transparent in HLET. Since the soft expansion is already performed at the Lagrangian level in the effective theory, the Dirac and propagator structure is significantly simpler in HLET. This, in turn, allows for the derivation of the  eikonal identities of Section~\ref{sec:eikid}, which form the basis of the above tree-level derivation and its all-order generalisation presented in the following section.

\subsection{$n$-photon emission to all orders} \label{sec:hlet_nphoton_allorder}

In what follows, the tree-level result of the previous section is generalised to all orders, which completes the proof of the soft theorem~\eqref{eq:lbk_ng_allorder}. In order to do so, virtual corrections are added to the generic diagrams~\eqref{eq:hlet_ngLP_general_tree}, \eqref{eq:hlet_ngNLPint_general_tree}, \eqref{eq:hlet_ngNLPext_general_tree}, and \eqref{eq:hlet_ngNLP_general_tree}. Due to the simple dependence on the loop momenta in the off-shell eikonal identity~\eqref{eq:eikid_gen}, the derivation of the previous section does not change substantially. In particular, the amplitudes reduce to all-order single-emission contributions -- which have been shown to be one-loop exact in~\cite{Engel:2023ifn} -- and tree-level exact double-emission terms.

To all orders, the generic LP amplitude~\eqref{eq:hlet_ngLP_general_tree} is given by
\begin{align} \label{eq:hlet_ngLP_general_allorder_diag}
    \mathcal{A}_{k_1...k_n,\mathcal{O}_\text{LP}}^{i...i}
    &= \sum_{u,s,t} \begin{tikzpicture}[scale=.8,baseline={(0,-.1)}]
        	   \input{figures/hlet_ngLP_general_allorder}
       	   \end{tikzpicture} \, ,
\end{align}
where the sums go from zero to infinity. The dashed lines represent multiple photons with the given number. In addition to the dashed line for the $n$ external photons, there are three types of virtual corrections in~\eqref{eq:hlet_ngLP_general_allorder_diag}. Loops that only attach to the emitting leg labelled by $v_i$ are represented by the dashed line labelled by $s$. 
The remaining two types correspond to corrections that connect the emitting leg with the remaining ones and others that only correct the non-emitting legs. These are represented by the dashed lines labelled by $u$ and $t$, respectively. As a consequence of the on-shell eikonal identity~\eqref{eq:eikid_conv}, all loop corrections in~\eqref{eq:hlet_ngLP_general_allorder_diag} turn out to be scaleless. There is, however, a subtlety in interpreting~\eqref{eq:hlet_ngLP_general_allorder_diag} in the case of single-leg corrections. To avoid that some permutations of the photon lines result in external self-energy corrections, it is useful to first discuss the loops that connect different legs and then generalise the argument to the remaining corrections.

In the case of the corrections that connect different legs, the on-shell eikonal identity~\eqref{eq:eikid_conv} can be applied to write the amplitude as
\begin{align} \label{eq:hlet_ngLP_general_allorder_reduced_easier}
    \begin{tikzpicture}[scale=.8,baseline={(0,-.1)}]  	   \input{figures/hlet_ngLP_general_allorder_simpler}
       	   \end{tikzpicture}
    =  \prod_{a=1}^{n} \eik_i(k_a)  \frac{1}{u!} \prod_{l=1}^u \int [\mathrm{d}\ell_l]
    \frac{-Q_i v_{i,\mu_l}}{[\ell_l^2] [\ell_l \cdot v_i]} \Gamma_i^{\mu_1...\mu_u} u_{v_i}  \, .
\end{align}
The combinatorial factor $u!$ takes into account the double counting due to the photon lines connecting two crossed vertices. The $d$-dimensional loop measure
\begin{align}
    [\mathrm{d}\ell] = i \mu^{2\epsilon} \frac{\mathrm{d}^d \ell}{(2\pi)^d}
\end{align}
is defined such that it absorbs the $i$ from the corresponding virtual photon propagator. The sub-amplitude $\Gamma_i^{\mu_1...\mu_u}$ represents the remaining part of the amplitude. Upon using the on-shell eikonal identity~\eqref{eq:eikid_conv} it takes the form
\begin{align}
    \Gamma_i^{\mu_1...\mu_u}
    = \sum_{j_1,...,j_u \neq i} 
    \prod_{l=1}^u \frac{Q_{j_l} v_{j_l}^{\mu_l}}{[-\ell_l \cdot v_{j_l}]} \Gamma_i \, ,
\end{align}
with $\mathcal{A}_m = \Gamma_i u_{v_i}$. All integrals in~\eqref{eq:hlet_ngLP_general_allorder_reduced_easier} are therefore scaleless. The additional loops in~\eqref{eq:hlet_ngLP_general_allorder_diag} labelled by $s$ and $t$ are now correctly taken into account by inserting one-particle irreducible (1PI) corrections at the crossed vertices. In this way, the aforementioned problem of external self-energy corrections is avoided. After summing over all attachments to the 1PI diagram, all momenta of the external photons completely decouple and the 1PI insertions are rendered scaleless. The diagram on the r.h.s.\ of~\eqref{eq:hlet_ngLP_general_allorder_diag} thus collapses to the tree-level amplitude~\eqref{eq:nonrad_matching} with $u=s=t=0$ and evaluates to
\begin{align} \label{eq:hlet_ngLP_general_allorder}
    \mathcal{A}_{k_1...k_n,\mathcal{O}_\text{LP}}^{i...i}
    = \prod_{a=1}^{n} \eik_i(k_a) \mathcal{A}_{\mathcal{O}_\text{LP}}
    = \prod_{a=1}^{n} \eik_i(k_a) \mathcal{A}_m \, .
\end{align}
The full all-order form of the matching relation~\eqref{eq:nonrad_matching} is used here for the first time.

In complete analogy to the tree-level case given in~\eqref{eq:hlet_ngNLPint_general_tree} and \eqref{eq:hlet_ngNLPext_general_tree}, the all-order $\mathcal{O}_\text{NLP}$ diagrams can be written in terms of the corresponding all-order single-emission amplitudes as
\begin{align}
    \label{eq:hlet_ngNLP_int_general_allorder}
    \mathcal{A}_{k_1...k_{n-1};k_n,\mathcal{O}_\text{NLP}}^{i...i}
    &= \sum_{u,s,t} \begin{tikzpicture}[scale=.8,baseline={(0,-.1)}]
        	   \input{figures/hlet_ngNLP_int_general_allorder}
       	   \end{tikzpicture}
    = \prod_{c=1}^{n-1} \eik_i(k_c) \mathcal{A}_{;k_n,\mathcal{O}_\text{NLP}}^{i} \, ,
    \\
    \label{eq:hlet_ngNLP_ext_general_allorder}
    \mathcal{A}_{k_1...k_n,\mathcal{O}_\text{NLP}}^{i...i}
    &= \sum_{u,s,t} \begin{tikzpicture}[scale=.8,baseline={(0,-.1)}]
        	   \input{figures/hlet_ngNLP_ext_general_allorder}
       	   \end{tikzpicture}
    = \sum_{a=1}^n\ \prod_{c \neq a} \eik_i(k_c) \mathcal{A}_{k_a,\mathcal{O}_\text{NLP}}^{i}\, .
\end{align}
In principle, an additional diagram type exists where a virtual photon connects to the $\mathcal{O}_\text{NLP}$ vertex instead of the external one in~\eqref{eq:hlet_ngNLP_int_general_allorder}. However, as a consequence of the on-shell eikonal identity~\eqref{eq:eikid_conv} the corresponding loop integral completely factorises and is rendered scaleless. Hence, these diagrams vanish and do not have to be considered explicitly. Furthermore, the all-order single-emission amplitudes $\mathcal{A}_{;k_n,\mathcal{O}_\text{NLP}}$ and $\mathcal{A}_{k_a,\mathcal{O}_\text{NLP}}^{i}$ have already been considered in~\cite{Engel:2023ifn}. In particular, it was shown that their contribution to the unpolarised squared amplitude can be computed via the naive all-order extension of the matching equation~\eqref{eq:relation_hlet_qed_tree} given by
\begin{align} \label{eq:relation_hlet_qed_allorder}
    \sum_\text{pol} \Big|
    \Big( \sum_i \mathcal{A}_{k_a,\mathcal{O}_\text{LP}}^{i}
    + \mathcal{A}_{k_a,\mathcal{O}_\text{NLP}}^{i} \Big)
    + \mathcal{A}_{;k_a,\mathcal{O}_\text{NLP}} \Big|^2
    + \mathcal{O}(\lambda^0)
    = (E(k_a) + D(k_a)) \mathcal{M}_m + \mathcal{O}(\lambda^0) \, .
\end{align}
There are two reasons for this. First, all virtual corrections in~\eqref{eq:hlet_ngNLP_int_general_allorder} and \eqref{eq:hlet_ngNLP_ext_general_allorder} are scaleless and therefore vanish. This is completely analogous to the LP case discussed above. Second, the kinetic vertices only contribute at one loop (and not at tree level) and the magnetic contribution vanishes completely at the level of the unpolarised squared amplitude. Hence, the l.h.s.\ of \eqref{eq:relation_hlet_qed_allorder} makes up the total purely hard contribution in HLET. It can therefore be calculated by expanding the QED amplitude with the method of regions in the hard momentum region, \textit{i.e.}\ in the region where all loop momenta scale as $\sim \lambda^0$. It is, however, precisely these purely hard contributions where the original tree-level proof of the LBK theorem presented in Section~\ref{sec:qed_1photon} still applies. This is because  internal emission from hard loops does not break locality. The only exception are factorisable corrections of the emitting leg. As shown in Section 5 of~\cite{Engel:2023ifn} these corrections are in one-to-one correspondence with the magnetic contribution in HLET and thus vanish for unpolarised scattering.

The all-order generalisation of the generic NLP diagram~\eqref{eq:hlet_ngNLP_general_tree} is given by
\begin{align} \label{eq:hlet_ngNLP_general_allorder}
    \mathcal{A}^{nuts,i}_{pwrq,V}
    &= \sum_{K \subset N} \begin{tikzpicture}[scale=.8,baseline={(0,-.1)}]
        	   \input{figures/hlet_ngNLP_general_allorder}
       	   \end{tikzpicture}
    + (k_{1,...,p} \leftrightarrow k_{p+1,...,n}) \, ,
\end{align}
where we omit all dashed lines that directly result in scaleless integrals after applying the on-shell eikonal identity~\eqref{eq:eikid_conv}. The index sets $K$, $P$ and $\bar{K}$ are defined as in~\eqref{eq:hlet_ngNLP_general_tree}. Furthermore, the virtual dashed lines are given in pairs with a fixed combined number of photons, \textit{e.g.}\ $w+(u-w)=u$. This yields a decomposition of the amplitude into gauge-invariant subsets at each given order in perturbation theory. The labels for the loop momenta decompose into the sets $\{\ell_1,...\ell_{w}\}$, $\{\ell_{w+1},...\ell_{u}\}$, $\{\ell_{u+1},...,\ell_{u+s-q}\}$ for photon lines that connect different external legs and $\{\tilde{\ell}_1,...\tilde{\ell}_{r}\}$, $\{\tilde{\ell}_{r+1},...\tilde{\ell}_t\}$, $\{\tilde{\ell}_{t+1},...\tilde{\ell}_{t+q}\}$ for the ones that only correct the leg labelled by $v_i$. The all-order magnetic and kinetic contributions can be constructed based on the generic amplitude~\eqref{eq:hlet_ngNLP_general_allorder} via
\begin{align}
    \mathcal{A}_{k_1...k_n,\text{mag}}^{i...i} \label{eq:amps_nlp_correspondence_mag}
    &= \sum_{u,t} \sum_{w=0}^{u} \sum_{r=0}^{t}
    \Big( \mathcal{A}^{nut0,i}_{1wr0,\text{mag}}
    + \mathcal{A}^{nut1,i}_{0wr0,\text{mag}}
    + \mathcal{A}^{nut1,i}_{0wr1,\text{mag}}
    \Big) \, ,
    \\
    \mathcal{A}_{k_1...k_n,0\gamma}^{i...i} \label{eq:amps_nlp_correspondence_kin0g}
    &= \sum_{u,t} \sum_{w=0}^{u} \sum_{r=0}^{t}
    \mathcal{A}^{nut0,i}_{0wr0,0\gamma} \, ,
    \\
    \mathcal{A}_{k_1...k_n,1\gamma}^{i...i} \label{eq:amps_nlp_correspondence_kin1g}
    &= \sum_{u,t} \sum_{w=0}^{u} \sum_{r=0}^{t}
    \Big( \mathcal{A}^{nut0,i}_{1wr0,1\gamma}
    + \mathcal{A}^{nut1,i}_{0wr0,1\gamma}
    + \mathcal{A}^{nut1,i}_{0wr1,1\gamma}
    \Big) \, ,
    \\
    \mathcal{A}_{k_1...k_n,2\gamma}^{i...i} \label{eq:amps_nlp_correspondence_kin2g}
    &= \sum_{u,t} \sum_{w=0}^{u} \sum_{r=0}^{t}
    \Big( \mathcal{A}^{nut0,i}_{2wr0,2\gamma}
    + \mathcal{A}^{nut1,i}_{1wr0,2\gamma}
    + \mathcal{A}^{nut1,i}_{1wr1,2\gamma}
    + \mathcal{A}^{nut2,i}_{0wr0,2\gamma}
    + \mathcal{A}^{nut2,i}_{0wr1,2\gamma}
    + \mathcal{A}^{nut2,i}_{0wr2,2\gamma}
    \Big) \, .
\end{align}
As for the tree-level diagram~\eqref{eq:hlet_ngNLP_general_tree}, the on-shell eikonal identity~\eqref{eq:eikid_conv} can be used for the crossed vertices that attach to an external leg. For the crossed vertex that corrects the off-shell leg, the off-shell identity~\eqref{eq:eikid_gen} applies with
\begin{align}
    \widetilde{\sum}_j \ell_j \cdot v_i
	& =
	\sum_{j=w+1}^{u} \ell_j \cdot v_i + \sum_{j=1}^{t+q} \tilde{\ell}_j \cdot v_i \, ,
	\\
	\tilde{p}  \cdot v_i
	& =
    \sum_{j=1}^w \ell_j \cdot v_i
	+ \sum_{j=u+1}^{u+s-q} \ell_j \cdot v_i
    - \sum_{j \in K} k_j \cdot v_i - \sum_{j=1}^p k_j \cdot v_i \, .
\end{align}
As a result, the amplitude~\eqref{eq:hlet_ngNLP_general_allorder} takes the form
\begin{align} \label{eq:hlet_ngNLP_general_allorder_amp}
	\mathcal{A}_{pwrq,V}^{nuts,i} 
	&= \prod_{c=1}^p \epsilon_c^{\rho_c} \prod_{c=p+1}^n \eik_i(k_c)
    \prod_{l=1}^{u+s-q} \int [\mathrm{d} \ell_l]
    \prod_{l=1}^{t+q} \int [\mathrm{d} \tilde{\ell}_l] \,
    \Gamma_i^{\nu_{q+1}...\nu_s\mu_1...\mu_u}
	(-1)^s S^{wu} S^{rt}
    \nonumber \\ & \quad \times
    \sum_{K \subset N} 
    F_{\nu_1 ... \nu_s\rho_1 ... \rho_p,V}^{wrq,K}
    \sum_{J \subset \bar{K}} (-1)^{|J|}
    \mathcal{P}_{utsq,\mu_1 ... \mu_{u}}^{\nu_1 ... \nu_q,\tilde{J}} u_{v_i}
    + (k_{1,...,p} \leftrightarrow k_{p+1,...,n}) \, ,
\end{align}
with the propagator structure
\begin{align}
     \mathcal{P}_{utsq,\mu_1 ... \mu_u}^{\nu_1 ... 
     \nu_q,\tilde{J}}
	 = &\Bigg( \prod_{l=1}^{u}  \frac{Q_i v_{i,\mu_l}}{[\ell_l^2] [\ell_l \cdot v_i]} \Bigg)
	\Bigg( \prod_{l=u+1}^{u+s-q}  \frac{1}{[\ell_l^2]} \Bigg)
    \Bigg( \prod_{l=1}^{t} \frac{Q_i^2}{[\tilde{\ell}_l^2] [\tilde{\ell}_l \cdot v_i] [-\tilde{\ell}_l \cdot v_i]} \Bigg)
	\Bigg( \prod_{l=t+1}^{t+q} \frac{Q_i v_i^{\nu_i}}{[\tilde{\ell}_l^2] [-\tilde{\ell}_l \cdot v_i]}  \Bigg)
	 \nonumber \\ & \quad \times
	 \frac{i}{\sum_{j=1}^{u+s-q} \ell_j \cdot v_i + \sum_{j=1}^{t+q} \tilde{\ell}_j \cdot v_i 
    -\sum_{j \in \tilde{J}} k_j \cdot v_i - \sum_{j=1}^{p} k_j \cdot v_i} \, .
\end{align}
The sub-amplitude $\Gamma_i^{\nu_{q+1}...\nu_s\mu_1...\mu_u}$ denotes the remaining part of the amplitude. Since no NLP vertex attaches to this part of the diagram, the loop momenta factorise according to the on-shell eikonal identity. At tree level the sub-amplitude reduces to $\Gamma_i^{(0)}$ defined in~\eqref{eq:subamp_tree}.
Furthermore, there is a double counting of contributions in the cases where a dashed line connects two crossed vertices. The factor
\begin{align} \label{eq:double_counting_factor}
    S^{xy} = \frac{(-1)^x}{x!(y-x)!}
\end{align}
takes this into account for the two dashed lines with $x$ and $y-x$ photons. Furthermore, it also absorbs the corresponding signs coming from the off-shell eikonal identity~\eqref{eq:eikid_gen} and the $-1$ from the photon propagators. 

The vertex factor in~\eqref{eq:hlet_ngNLP_general_allorder_amp} for the magnetic contribution reads
\begin{align} \label{eq:vertexfac_allorder_mag}
    F_{\mu,\text{mag}}^{wrq,K}
    = \frac{Q_i C_{\text{mag}}}{4 m_i} (1+\slashed{v_i})\sigma_{\mu \lambda} p^\lambda \, ,
    \quad
    p^\lambda = \begin{cases}
        -k_1^\lambda, & q=0, \mu = \rho_1 \\
        \ell_{u+1}^\lambda, & q=0, \mu = \nu_1 \\
        \tilde{\ell}_{t+1}^\lambda, & q=1, \mu = \nu_1
    \end{cases} \, .
\end{align}
In the case of the three kinetic vertices it is given by
\begin{align} \label{eq:vertexfac_allorder_kin0g}
    F_{0\gamma}^{wr0,K}
    &= \frac{i}{2 m_i} 
    \Big(L -\sum_{j \in K} k_j \Big)_{\perp_i}^2 \, ,
    \\
     F_{\mu,1\gamma}^{wrq,K}
    &= \frac{-i Q_i}{2 m_i} 
    \Big(p + 2 L - 2 \sum_{j \in K} k_j \Big)^{\perp_i}_{\mu} \, ,
    \\
     F_{\mu_1 \mu_2,2\gamma}^{wrq,K}
    &= \frac{i Q_i^2}{m_i} g^{\perp_i}_{\mu_1 \mu_2} \, ,
\end{align}
with
\begin{align}
    L^\mu = \sum_{l=1}^w \ell_l^\mu + \sum_{l=1}^r \tilde{\ell}_l^\mu
\end{align}
and $p^\mu$ defined as in~\eqref{eq:vertexfac_allorder_mag}.

In complete analogy to the tree-level calculation of the previous section, the expression~\eqref{eq:hlet_ngNLP_general_allorder_amp} can be collected in terms of the propagator structures by reshuffling the sums according to~\eqref{eq:reshuffle_id}. This yields
\begin{align}   \label{eq:hlet_ngNLP_general_allorder_amp_reshuffled}
	\mathcal{A}_{pwrq,V}^{nuts,i} 
	&= \prod_{c=1}^p \epsilon_c^{\rho_c} \prod_{c=p+1}^n \eik_i(k_c)
    \prod_{l=1}^{u+s-q} \int [\mathrm{d} \ell_l]
    \prod_{l=1}^{t+q} \int [\mathrm{d} \tilde{\ell}_l] \,
    \Gamma_i^{\nu_{q+1}...\nu_s\mu_1...\mu_u}
	(-1)^s S^{wu} S^{rt}
    \nonumber \\ & \quad \times
    \sum_{\tilde{J} \subset N} 
    \bar{F}_{\nu_1 ... \nu_s\rho_1 ... \rho_p,V}^{wrq,\tilde{J}}
    \mathcal{P}_{utsq,\mu_1 ... \mu_{u}}^{\nu_1 ... \nu_q,\tilde{J}} u_{v_i}
    + (k_{1,...,p} \leftrightarrow k_{p+1,...,n}) \, ,
\end{align}
with the propagator coefficients
\begin{align} \label{eq:propcoeff_allorder}
    \bar{F}_{\nu_1 ... \nu_s\rho_1 ... \rho_p,V}^{wrq,\tilde{J}}
    = (-1)^{|\tilde{J}|} \sum_{K \subset \tilde{J}}
    F_{\nu_1 ... \nu_s\rho_1 ... \rho_p,V}^{wrq,K} \, .
\end{align}
The evaluation of this coefficient does not significantly change compared to the tree-level version~\eqref{eq:propcoeff_tree}. Also in this case the coefficients for the four vertex types turn out to vanish for most choices of $\tilde{J}$ and read
\begin{align}
    \bar{F}_{\mu,\text{mag}}^{wrq,\tilde{J}} \label{eq:propcoeff_allorder_mag}
    &= \frac{Q_i C_{\text{mag}}}{4 m_i} (1+\slashed{v_i})\sigma_{\mu \lambda} p^\lambda
    \begin{cases}
        1, & \tilde{J}=\{\} \\
        0, & \text{otherwise}
    \end{cases} \, ,
    \\
    \bar{F}_{0\gamma}^{wr0,\tilde{J}} \label{eq:propcoeff_allorder_0g}
    &= \frac{i}{2 m_i}
    \begin{cases}
        L^2, & \tilde{J}=\{\} \\
        (k_a-2L)^{\perp_i} \cdot k_{a}^{\perp_i}, & \tilde{J} = \{k_a\} \\
        2 k_{a}^{\perp_i} \cdot k_{b}^{\perp_i}, & \tilde{J} = \{k_a,k_b\} \\
        0, & \text{otherwise}
    \end{cases} \, ,
    \\
    \bar{F}_{\mu,1\gamma}^{wrq,\tilde{J}} \label{eq:propcoeff_allorder_1g}
    &= \frac{iQ_i}{2 m_i} 
    \begin{cases}
        (-p-2L)_{\mu}^{\perp_i}, & \tilde{J}=\{\} \\
        2 k_{a,\mu}^{\perp_i}, & \tilde{J} = \{k_a\} \\
        0, & \text{otherwise}
    \end{cases} \, ,
    \\
    \bar{F}_{\mu_1 \mu_2,2\gamma}^{wrq,\tilde{J}} \label{eq:propcoeff_allorder_2g}
    &= \frac{i Q_i^2}{m_i} g^{\perp_i}_{\mu_1 \mu_2}
    \begin{cases}
        1, & \tilde{J}=\{\} \\
        0, & \text{otherwise}
    \end{cases} \, .
\end{align}
The momentum $p^\mu$ in~\eqref{eq:propcoeff_allorder_mag} and~\eqref{eq:propcoeff_allorder_1g} is chosen according to~\eqref{eq:vertexfac_allorder_mag}. The above expressions reduce to the tree-level coefficients~\eqref{eq:vertexfac_tree_mag}-\eqref{eq:vertexfac_tree_kin2g} for $L=0$ and $q=0$. Furthermore, if $\tilde{J}=\{\}$ all loop integrals are trivially scaleless and we can set $L=0$ in this case. In particular, this implies that $\bar{F}_{0\gamma}^{wr0,\{\}}=0$. The above coefficients thus have exactly the same structure as the tree-level ones. Hence, also the all-order amplitude~\eqref{eq:hlet_ngNLP_general_allorder_amp_reshuffled} reduces to single- and double emission terms according to
\begin{align} 
    \label{eq:amp_allorder_reduced_mag}
     \mathcal{A}_{k_1...k_n,\text{mag}}^{i...i}
    &= \sum_{a=1}^n \prod_{c \neq a} \eik_i(k_c) \mathcal{A}_{k_a,\text{mag}}^{i} \, ,
    \\
    \label{eq:amp_allorder_reduced_kin}
    \mathcal{A}_{k_1...k_n,V}^{i...i}
    &= \sum_{a=1}^n \prod_{c \neq a} \eik_i(k_c) \mathcal{A}_{k_a,V}^{i}
    + \sum_{a,b=1}^n \prod_{c \neq a,b} \eik_i(k_c) \mathcal{G}_{i,V}(k_a,k_b) \mathcal{A}_m \, ,
\end{align}
with $V \in \{0\gamma,1\gamma,2\gamma\}$. This is the all-order generalisation of the tree-level relations~\eqref{eq:hlet_ng_tree_mag} and~\eqref{eq:hlet_ng_tree_magkin}. In~\cite{Engel:2023ifn}, the all-order single-emission amplitudes $\mathcal{A}_{k_a,\text{mag}}^{i}$ and $\mathcal{A}_{k_a,V}^{i}$ have been shown to be tree-level and one-loop exact, respectively. The corresponding argument is repeated in the following and used to prove that the all-order double-emission objects $\mathcal{G}_{i,V}(k_a,k_b)$ are tree-level exact.

Since $\bar{F}_{0\gamma}^{wr0,\{\}}=0$ as argued above, the coefficients~\eqref{eq:propcoeff_allorder_mag}-\eqref{eq:propcoeff_allorder_2g} are either independent or linearly dependent on $L$. In addition, they enter the expression~\eqref{eq:hlet_ngNLP_general_allorder_amp_reshuffled} that is completely symmetric under permutations of $\{\ell_{1},...,\ell_w\}$ and $\{\tilde{\ell}_{1},...,\tilde{\ell}_r\}$, which allows for the simplifying replacement
\begin{align}
    L^\mu = \sum_{l=1}^w \ell_l^\mu + \sum_{l=1}^r \tilde{\ell}_l^\mu
      \to w \ell_1^\mu + r \tilde{\ell}_1^\mu \, .
\end{align}
This, in turn, allows for the evaluation of the corresponding sums in~\eqref{eq:amps_nlp_correspondence_mag}-\eqref{eq:amps_nlp_correspondence_kin2g} with the identities
\begin{align}
    \sum_{x=0}^y S^{xy} &= \delta_{y,0} \, , \label{eq:loopsum1}
    \\
    \sum_{x=0}^y S^{xy} x &= -\delta_{y,1} \, . \label{eq:loopsum2}
\end{align}
The vertex factors that are independent of $w$ and $r$ force the loop corrections related to $u$ and $t$ to vanish. A linear dependence, on the other hand, gives rise to a one-loop exact contribution.

Genuine double-emission terms contributing to $\mathcal{G}_{i,V}$ in~\eqref{eq:amp_allorder_reduced_kin} only arise from the vertex factors $\bar{F}_{0\gamma}^{wr0,\{k_a,k_b\}}$, $\bar{F}_{\rho_1,1\gamma}^{wr0,\{k_a\}}$, and $\bar{F}_{\rho_1\rho_2,2\gamma}^{wr0,\{\}}$. All of them are independent of $w$ and $r$ which implies together with~\eqref{eq:loopsum1} that $u=0$ and $t=0$. Since in addition $q=0$, there are no loop corrections at all in this case. The genuine double-emission term in~\eqref{eq:amp_allorder_reduced_kin} is thus tree-level exact $\mathcal{G}_{i,V}=\mathcal{G}_{i,V}^{(0)}$ and the total kinetic amplitude simplifies to
\begin{align} \label{eq:amp_kin_2g_allorder}
     \mathcal{A}_{k_1...k_n,\text{kin}}^{i...i}
    = \sum_{a=1}^n \prod_{c \neq a} \eik_i(k_c) \mathcal{A}_{k_a,\text{kin}}^{i}
    + \sum_{a,b=1}^n \prod_{c \neq a,b} \eik_i(k_c) \mathcal{G}^{(0)}_{i}(k_a,k_b) \mathcal{A}_m\, .
\end{align}
Contrary to the tree-level case~\eqref{eq:hlet_1g_kin_total}, the total kinetic single-emission amplitude $\mathcal{A}_{k_a,\text{kin}}^{i}$ does not vanish. Instead, it gives rise to a one-loop exact contribution as shown in~\cite{Engel:2023ifn}. In the case of the magnetic amplitude $\mathcal{A}_{k_a,\text{mag}}^{i}$, on the other hand, all loop corrections vanish as for $\mathcal{G}_{i,V}$. In the following, the explicit results from~\cite{Engel:2023ifn} are repeated for the sake of completeness. 

The tree-level exact magnetic contribution is given by
\begin{align} \label{eq:amp_mag_ng_allorder}
    \mathcal{A}_{k_a,\text{mag}}^{i}
    = \frac{Q_i C_\text{mag}}{2 m_i} \Gamma_i \epsilon_a \cdot H_a u_{v_i} \, ,
\end{align}
with $H_a^\mu$ defined in~\eqref{eq:Htensor}. As already mentioned, the specific form of this tensor results in a vanishing contribution for unpolarised scattering. The result~\eqref{eq:amp_mag_ng_allorder} together with~\eqref{eq:amp_allorder_reduced_mag} therefore proves that the magnetic contribution does not enter the soft theorem~\eqref{eq:lbk_ng_allorder}. The result for the kinetic contribution is given in~(4.40) of~\cite{Engel:2023ifn} and reads
\begin{align} \label{eq:amp_kin_1g_allorder}
    \mathcal{A}_{k_a,\text{kin}}^{i}
    = \sum_{j \neq i} Q_i^2 Q_j \Big( \frac{\epsilon_a \cdot p_i}{k_a \cdot p_i}
    - \frac{\epsilon_a \cdot p_j§}{k_a \cdot p_j} \Big)
    \mathcal{S}^{(1)}(p_i,p_j,k_a) \mathcal{A}_m \, ,
\end{align}
with the one-loop exact soft function $\mathcal{S}^{(1)}$ defined in~\eqref{eq:softfunc}.

In summary, the all-order amplitude to NLP for an arbitrary number of soft-photon emissions takes the remarkably simple form
\begin{align} \label{eq:lbk_ng_allorder_amp}
    \mathcal{A}_{m+n}
    = \prod_{a=1}^n \eik(k_a) \mathcal{A}_m
    + \sum_{a=1}^n \prod_{c \neq a} \eik(k_c) \mathcal{A}_{k_a,\text{NLP}}
    + \sum_{a,b=1}^n \prod_{c \neq a,b} \eik(k_c) \mathcal{G}^{(0)}(k_a,k_b) \mathcal{A}_m
    + \mathcal{O}(\lambda^{-n+2}) \, ,
\end{align}
with the NLP single-emission contribution
\begin{align} \label{eq:single_nlp_hlet_allorder}
     \mathcal{A}_{k_a,\text{NLP}}
     = \Big( \sum_i \mathcal{A}_{k_a,\mathcal{O}_\text{NLP}}^{i} 
        + \mathcal{A}_{k_a,\text{mag}}^{i}
        + \mathcal{A}_{k_a,\text{kin}}^{i} \Big)
        + \mathcal{A}_{;k_a,\mathcal{O}_\text{NLP}} \, .
\end{align}
The results for $\mathcal{A}_{k_a,\text{mag}}^{i}$ and $\mathcal{A}_{k_a,\text{kin}}^{i}$ are given in~\eqref{eq:amp_mag_ng_allorder} and~\eqref{eq:amp_kin_1g_allorder}, respectively. The magnetic contribution vanishes in the unpolarised case. The tree-level exact genuine double-emission contribution $\mathcal{G}^{(0)}$ is defined in~\eqref{eq:eik_sum}. Finally, the contribution of the $\mathcal{O}_\text{NLP}$ amplitudes in~\eqref{eq:single_nlp_hlet_allorder} to the unpolarised squared amplitude is given by the matching relation~\eqref{eq:relation_hlet_qed_allorder}. This reproduces the soft theorem~\eqref{eq:lbk_ng_allorder} and completes the all-order proof.

\section{Real-real-virtual corrections for muon-electron scattering} \label{sec:validation}

This section considers muon-electron scattering as a first non-trivial application of the NLP soft theorem~\eqref{eq:lbk_ng_allorder}. This process has gained considerable attention in recent years~\cite{Banerjee:2020rww,Broggio:2022htr,CarloniCalame:2020yoz,Alacevich:2018vez,Budassi:2021twh,Budassi:2022kqs,Fael:2019nsf,Fael:2018dmz,Masiero:2020vxk,Dev:2020drf,Schubert:2019nwm,GrillidiCortona:2022kbq,Galon:2022xcl,Asai:2021wzx,Badger:2023xtl} due to the MUonE experiment~\cite{CarloniCalame:2015obs,Abbiendi:2016xup,Spedicato:2022Nb,Abbiendi:2022oks} requiring a high-precision theory prediction at the level of 10 parts per million (ppm). The MUonE experiment aims at extracting the hadronic contribution to the running of the QED coupling $\alpha$ with a precision below $1\%$. Currently, the two independent Monte Carlo codes MESMER~\cite{CarloniCalame:2020yoz} and {\sc McMule}~\cite{Banerjee:2020rww} incorporate NNLO QED corrections for muon-electron scattering. Both codes calculate the contributions that only correct the electron and muon line without any approximation. In~\cite{CarloniCalame:2020yoz}, the genuine two-loop four-point topologies are approximated with a YFS-inspired approach. The calculation presented in~\cite{Broggio:2022htr}, on the other hand, takes these into account by massifying~\cite{Penin:2005eh,Mitov:2006xs,Becher:2007cu,Engel:2018fsb} the amplitude with a vanishing electron mass~\cite{Mastrolia:2017pfy,DiVita:2018nnh,Bonciani:2021okt,Mandal:2022vju}. This gives a correct description up to terms that are polynomially suppressed by the electron mass. 

The differential results presented in~\cite{Broggio:2022htr} indicate that corrections beyond NNLO are required to meet the 10ppm precision goal. This has triggered a collaborative effort to calculate the dominant electron-line corrections to muon-electron scattering at N$^3$LO~\cite{Durham:n3lo}. Two major steps in this direction have already been taken with the calculation of the heavy-quark form factor at three loops~\cite{Fael:2022rgm,Fael:2022miw,Fael:2023zqr} and the construction of the all-order FKS$^\ell$ subtraction scheme~\cite{Engel:2019nfw}. One of the remaining challenges is the numerical stability of the real-real-virtual amplitude. In the case of real-virtual corrections at NNLO, the method of next-to-soft stabilisation~\cite{Banerjee:2021mty} has proven to ensure stable evaluations for a wide range of processes~\cite{Banerjee:2021mty,Banerjee:2021qvi,Engel:2023arz} and in particular also for muon-electron scattering~\cite{Broggio:2022htr}. Most numerical instabilities in real-emission amplitudes occur in the region of soft-collinear photon emission. By replacing the full amplitude in this delicate region with its soft NLP expansion, an efficient, stable, and accurate implementation is obtained. In addition, the LBK theorem~\eqref{eq:lbk_allorder} provides a fully automatable  way to obtain the corresponding expansion based on lower-order amplitudes. The generalisation~\eqref{eq:lbk_ng_allorder} of the LBK theorem to multi-photon emission therefore provides the basis to extend next-to-soft stabilisation to multiple radiation.

In the following, the one-loop electron line corrections to the process
\begin{align}
    e^{-}(p_1) \mu^{-}(p_2) \to e^{-}(p_3) \mu^{-}(p_4) \gamma(k_1) \gamma(k_2)
\end{align}
are calculated at NLP in the unordered soft limit $k_1 \sim k_2 \ll p_i,m_i$. The soft theorem~\eqref{eq:lbk_ng_allorder} is used with the charge and momentum signs
\begin{equation}
\begin{alignedat}{10} 
\label{eq:signs}
&p_1&\to&+p_1,&\qquad   &p_2&\to&+p_2,&\qquad
&p_3&\to&-p_3,&\qquad   &p_4&\to&-p_4\, ,&\qquad
\\
&Q_1&=  &- e,&   &Q_2&=  &- e,&
&Q_3&=  &+ e,&   &Q_4&=  &+ e \, .
\end{alignedat}
\end{equation}
These sign conventions also have to be taken into account in the derivatives $\partial/\partial p_i^\mu$ of~\eqref{eq:lbk_inv}. Furthermore, the non-radiative invariants are chosen as
\begin{align}
	\label{eq:invariants}
    \{s\}=\{s=(p_1+p_2)^2,\,
            t=(p_2-p_4)^2 \} \, .
\end{align}
This definition has to be consistently used both in the evaluation of the non-radiative amplitude as well as in the calculation of the derivatives $\partial s_L/\partial p_i^\mu$. The LBK operators~\eqref{eq:lbk_inv} for the incoming and outgoing electron then read
\begin{align}
	\tilde{\mathcal{D}}_1^\mu(k_a)
	&= \Big(\frac{p_1^\mu}{k_a \cdot p_1} k_a \cdot p_2 - p_2^\mu \Big) 2 \frac{\partial}{\partial s}\, , 
    \\
	\tilde{\mathcal{D}}_3^\mu(k_a)
	&= 0 \, .
\end{align}
The operators $\tilde{\mathcal{D}}_2^\mu$ and $\tilde{\mathcal{D}}_4^\mu$ correspond to the muon line and do not enter the electron-line corrections. The only remaining input needed for the soft theorem~\eqref{eq:lbk_ng_allorder} is the non-radiative one-loop amplitude which can easily be calculated based on the heavy-quark form factor~\cite{Mastrolia:2003yz,Bonciani:2003ai,Bernreuther:2004ih,Gluza:2009yy}. The corresponding result is expressed in terms of harmonic polylogarithms~\cite{Remiddi:1999ew} that can be evaluated with the \texttt{Mathematica} code \texttt{HPL}~\cite{Maitre:2005uu,Maitre:2007kp}.

\begin{figure}
    \centering
    \subfloat[pole]{
        \includegraphics[width=.7\textwidth]{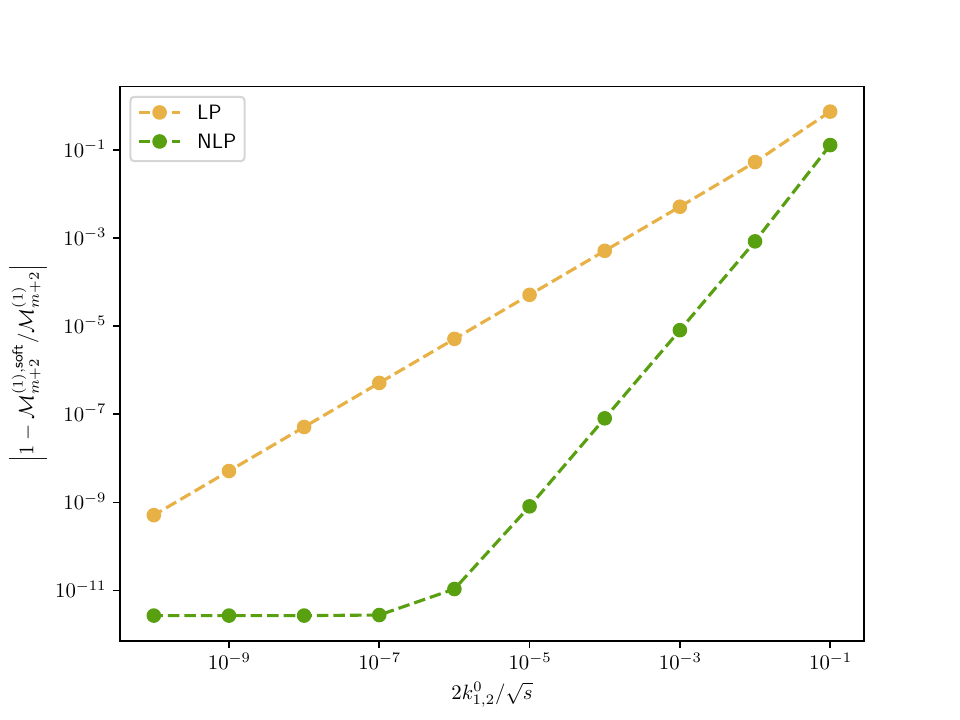}
    \label{fig:pole}
    } \\
    \subfloat[finite]{
        \includegraphics[width=.7\textwidth]{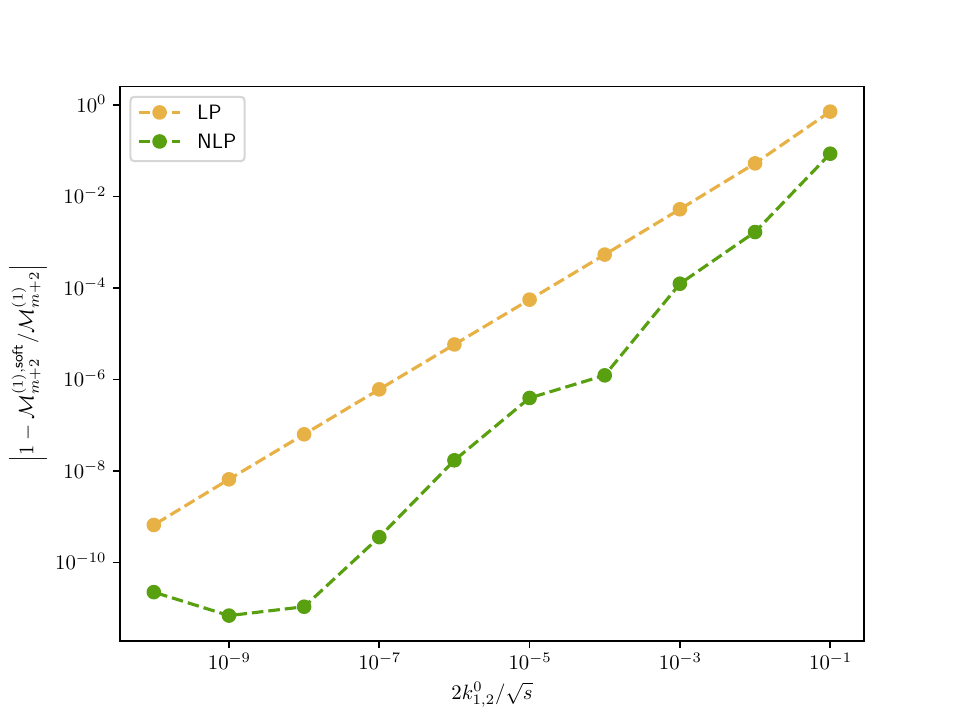}
    \label{fig:finite}
    }
\caption{The convergence of the soft approximation at LP and NLP as a function of the normalised photon energies $2 k_{1,2}^0/\sqrt{s}$. The exact reference values have been calculated using {\sc OpenLoops} in quadruple-precision mode. As expected, the relative difference of the soft approximation to the exact value tends to zero for decreasing photon energies $k_{1,2}^0 \to 0$. The inclusion of the NLP term gives a significant improvement of the approximation both for the IR pole as well as for the finite term. Beyond the relative precision $10^{-11}$ no further improvement is observed due to the finite precision of the exact computation.}
\label{fig:muone}
\end{figure}

As a verification, the NLP expansion obtained in this way is compared with the exact computation of the amplitude. The YFS exponentiation formula~\cite{Yennie:1961ad}
\begin{equation}
	e^{\hat{\mathcal{E}}} \mathcal{M}_{m+2} 
	= e^{\hat{\mathcal{E}}}  \sum_{\ell=0}^\infty \mathcal{M}_{m+2}^{(\ell)}
	= \text{finite}
\end{equation}
yields for the exact IR pole
\begin{equation}
	\mathcal{M}_{m+2}^{(1)}
	= - \hat{\mathcal{E}} \mathcal{M}_{m+2}^{(0)}
	+ \mathcal{O}(\epsilon^0) \, .
\end{equation}
The explicit form of the integrated eikonal $\hat{\mathcal{E}}$ is given in~\cite{Frederix:2009yq}. To compute the double-radiative tree-level amplitude, $\mathcal{M}_{m+2}^{(0)}$, \texttt{QGraf}~\cite{NOGUEIRA1993279} was used to generate the corresponding diagrams and \texttt{Package-X}~\cite{Patel:2015tea} to evaluate them. After expanding the result in the soft momentum to NLP, analytic agreement is found with the prediction from the NLP soft theorem. An analytic comparison of the finite part, on the other hand, would be much more involved. Instead a numerical comparison is performed with {\sc OpenLoops}~\cite{Buccioni:2017yxi,Buccioni:2019sur} running in quadruple precision. Figure~\ref{fig:muone} shows the relative difference between the LP and NLP soft approximations as a function of the normalised photon energies $2k_{1,2}^0/\sqrt{s}$. For completeness the convergence is shown both for the IR pole (Figure~\ref{fig:pole}) as well as for the finite part (Figure~\ref{fig:finite}). In both cases the inclusion of the NLP term in the soft expansion results in a significant improvement of the approximation. Due to the finite precision of the exact reference values, the convergence saturates at the level of $10^{-11}$. Figure~\ref{fig:muone} thus represents a non-trivial validation of the NLP soft theorem~\eqref{eq:lbk_ng_allorder}.

\section{Conclusion} \label{sec:conclusion}

This paper studied the behaviour of multi-photon emission amplitudes in the soft limit at NLP and to all orders. The simple structure in this limit was made transparent by working in HLET, the Abelian version of HQET. Following the methodology developed for single soft-photon radiation in~\cite{Engel:2023ifn}, a generalised off-shell eikonal identity for multiple emissions was derived, which forms the basis of the proof. Compared to the single-emission case, the presence of multiple soft scales results in fewer scaleless integrals and therefore in a more intricate structure of the identity. This, in turn, significantly complicates the combinatorics involved in the analysis of the all-order amplitudes. Nevertheless, it is possible to write the amplitudes in a form such that a simple structure emerges. Specifically, the NLP multi-emission amplitudes were shown to reduce to single- and double-emission contributions only. At this point, the approach of the single-emission study~\cite{Engel:2023ifn} can be applied to show that the genuine double-emission contribution is tree-level exact. Combined with the one-loop exact soft function for single emission, this results in the simple form of the NLP soft theorem given in~\eqref{eq:lbk_ng_allorder}. As a validation and a first non-trivial application of this theorem, the real-real-virtual electron-line corrections to muon-electron scattering were calculated at NLP in the soft limit.

The result of this article generalises the LP soft theorem of Yennie, Frautschi, and Suura to NLP. This opens the door for several interesting applications. First of all, it provides the basis for the application of next-to-soft-stabilisation to multi-emission amplitudes. This is particularly useful for local subtraction schemes, which require a stable evaluation of radiative amplitudes deep into the IR region. In the case of slicing schemes, on the other hand, it offers the opportunity to take into account power corrections, which, in turn, allows for larger values of the slicing cut. Furthermore, fixed-order computations could be supplemented with the effect of an arbitrary number of NLP soft-photon emissions, extending existing YFS frameworks for the resummation of soft logarithms. Finally, the theorem can be used to cross check exact amplitudes or, in the case where such a computation is not possible, to `bootstrap' the result.

\subsection*{Acknowledgement} 

I thank Mathieu Pellen, Marco Rocco, Maximilian Stahlhofen, and Yannick Ulrich for the careful reading of the manuscript as well as valuable suggestions for improvement. I was supported by the German Federal Ministry for
Education and Research (BMBF) under contract no.\ 05H21VFCAA.

\bibliographystyle{JHEP}
\bibliography{main.bib}

\end{document}